\documentclass[%
reprint,
amsmath,amssymb,
aps,
]{revtex4-1}

\usepackage{braket}
\usepackage{graphicx}
\usepackage{amsmath}
\usepackage{bm}
\usepackage{hhline}
\usepackage{natbib}

\usepackage{color}%
\definecolor{mygreen}{rgb}{0.0,0.7,0.0}

\begin{document}
\bibliographystyle{apsrev4-2}
%\preprint{}

\title{Anomalous dielectric response in insulators with the $\pi$ Zak phase}


%\author{Yisuke Aihara, Motoaki Hirayama and Shuishi Murakami}
\author{Yusuke Aihara}
\email{These authors contributed equally to this work.}
%\thanks{}
\affiliation{Department of Physics, Tokyo Institute of Technology,2-12-1 Ookayama, Meguro-ku, Tokyo 152-8551, Japan}
\author{Motoaki Hirayama}
\email{These authors contributed equally to this work.}
\affiliation{RIKEN Center for Emergent Matter Science, Wako, Saitama 351-0198, Japan}
\author{Shuichi Murakami}
\affiliation{Department of Physics, Tokyo Institute of Technology,2-12-1 Ookayama, Meguro-ku, Tokyo 152-8551, Japan}
\affiliation{TIES, Tokyo Institute of Technology, 2-12-1 Ookayama, Meguro-ku, Tokyo 152-8551, Japan}


\date{\today}

\begin{abstract}
In various topological phases, nontrivial states appear at the boundaries of the system. In this paper, we investigate anomalous dielectric response caused by such states caused by the $\pi$ Zak phase. First, by using the one-dimensional Su-Schrieffer-Heeger model, we show that, when the system is insulating and the Zak phase is $\pi$, the polarization suddenly rises to a large value close to $e/2$, by application of an external electric field. The $\pi$ Zak phase indicates existence of half-filled edge states, and we attribute this phenomenon to charge transfer between the edge states at the two ends of the system. We extend this idea to two- and three-dimensional insulators with the $\pi$ Zak phase over the Brillouin zone, and find similar anomalous dielectric response. We also show that diamond and silicon slabs with (111) surfaces have the $\pi$ Zak phase by {\it ab intio} calculations, and show that this anomalous response survives even surface reconstruction involving an odd number of original surface unit cells. Another material example with an anomalous dielectric response is polytetrafluoroethylene (PTFE), showing plateaus of polarization at $\pm e$ by {\it ab initio} calculation, in agreement with our theory. 
\end{abstract}

%\keywords{}

\maketitle

\section{introduction}
%%%%%%%%%%%%%%%%%%%%%%%%%%%%%%%%%%%%%%%%%%%%%%%%%%%%%
Topological materials such as topological insulators are currently under intense investigations due to their unique properties \cite{kane2005z,fu2007topological,bernevig2006quantum}. They are called topological because the $k$-space structure of electronic bands has nontrivial topology. In the topological phase, nontrivial localized states generally appear at the boundaries of the system, 
and this important correspondence is called bulk-edge correspondence~\cite{fu2006time,fu2007topological}. These edge/surface states often cause distinctive phenomena such as quantum spin Hall effect. 

In this paper we focus on topological phases characterized by the Zak phase. 
The Zak phase \cite{zak1989berry} is defined as a special case of the Berry phase \cite{berry1984quantal}. It is useful to determine presence or absence of topological edge states in systems with time-reversal (TR) and space inversion (SI) symmetries.
In spinless systems, the
Zak phase is quantized to be $0$ or $\pi$ under such symmetries and the 
systems have
topological edge states in the latter case \cite{zak1989berry,ryu2002topological,delplace2011zak,kariyado2013symmetry,Rhim,Miert,PhysRevB.101.161106}. The Zak phase is also known to represent the bulk polarization of the system
\cite{resta1992r,Vanderbilt1,Vanderbilt2,Resta2}, and at the same time the amount of surface charge on a crystalline solid \cite{Vanderbilt1,Vanderbilt2}. In systems with TR and SI symmetries, the surface charge per unit cell is quantized to be $0$ or $e/2$ modulo $e$ due to the quantization of the Zak phase where $-e$ is the electron charge. For one-dimensional systems, 
no edge charge accumulates at both ends when the Zak is 0 and the charge $e/2$ (mod $e$) accumulates at each edge when the Zak phase is $\pi$. For two- or three-dimensional insulating systems with TR and SI symmetries, in which the Zak phase is constant over the Brillouin zone, no edge charge per unit cell accumulates at both surfaces when the Zak is 0 and the charge $e/2$ (mod $e$) per unit cell accumulates at both surfaces when the Zak phase is $\pi$. These huge surface charges do not contradict the inversion symmetry because an equal amount of charges is accumulated at both ends. 
In half-filled one-dimensional insulating systems with the $\pi$ Zak phase, the edge states at both ends of the system are half-filled under SI symmetry. These edge states are expected to respond sensitively to an external field which breaks the symmetry.

In this paper, we apply a static electric field to such insulators with the $\pi$ Zak phase and calculate their dielectric response. As a result, we find a sudden uprise of the polarization even for a very weak electric field. First we show this phenomena in a one-dimensional 
Su-Schrieffer-Heeger (SSH) model with a Zak phase equal to $\pi$, which 
in turn has topological in-gap edge states of the chain. 
We also confirm that this anomalous phenomenon is attributed to the topological edge states. In two- and three-dimensional insulating systems with $\pi$ Zak phase, we find a similar anomalous behavior of the polarization when the edge or surface bands are sufficiently flat. 
Insulators with the $\pi$ Zak phases have actually been found, such as silicon
and diamond \cite{Vanderbilt2}, and some topological electrides \cite{hirayama2017topological,hirayama2018electrides}. For example, diamond and silicon slabs with the (111) surfaces have the $\pi$ Zak phase and relatively flat midgap states. 
 We also 
show such a dielectric response of polytetrafluoroethylene (PTFE)  with plateaus at $P\sim \pm e$ by {\it ab initio} calculation, in agreement with our theory. 
Our results are expected to be applicable to such materials.

The organization of the paper is as follows. In Sec.~\ref{SSH}, we first confirm that the anomalous dielectric response occurs in the one-dimensional SSH model and also confirm that the phenomenon is caused by the topological edge states. In Sec.~\ref{3D}, we also show similar phenomena in a three-dimensional model.
We discuss several points on realizing anomalous dielectric responses in Sec.~\ref{sec:remark}.
In Sec.~\ref{sec:material}, we show that diamond and silicon slabs with the (111) surfaces have the $\pi$ Zak phase leading to midgap surface states. We also show an anomalous dielectric response of polytetrafluoroethylene
(PTFE).
We then discuss the relationship between our results and the bulk polarization 
described in the modern theory of polarization in Sec.~\ref{sec:bulk-surface}.
We summarize our result in Sec. \ref{Conclusion}.
Throughout the paper, we consider systems with weak spin-orbit coupling, and
ignore it. Thus the polarizations and surface charges in the subsequent results should be doubled, in order 
to include the spin degree of freedom.

%%%%%%%%%%%%%%%%%%%%%%%%%%%%%%%%%%%%%%%%%%%%%%%%%%%
\section{One-dimensional model (Model I)}\label{SSH}
%%%%%%%%%%%%%%%%%%%%%%%%%%%%%%%%%%%%%%%%%%%%%%%%%%%
In this chapter, we show anomalous dielectric response in the SSH model \cite{su1979solitons}, which we call Model I. This model becomes a one-dimensional topological insulator by adjusting its parameters. This topological insulator phase is characterized by the value of the Zak phase equal to $\pi$ and this model has topological edge states. 
%%%
\subsection{Previous research}
%%%FIG%1%%
   \begin{figure}[tbp]
    \centering
    \includegraphics[width=86mm]{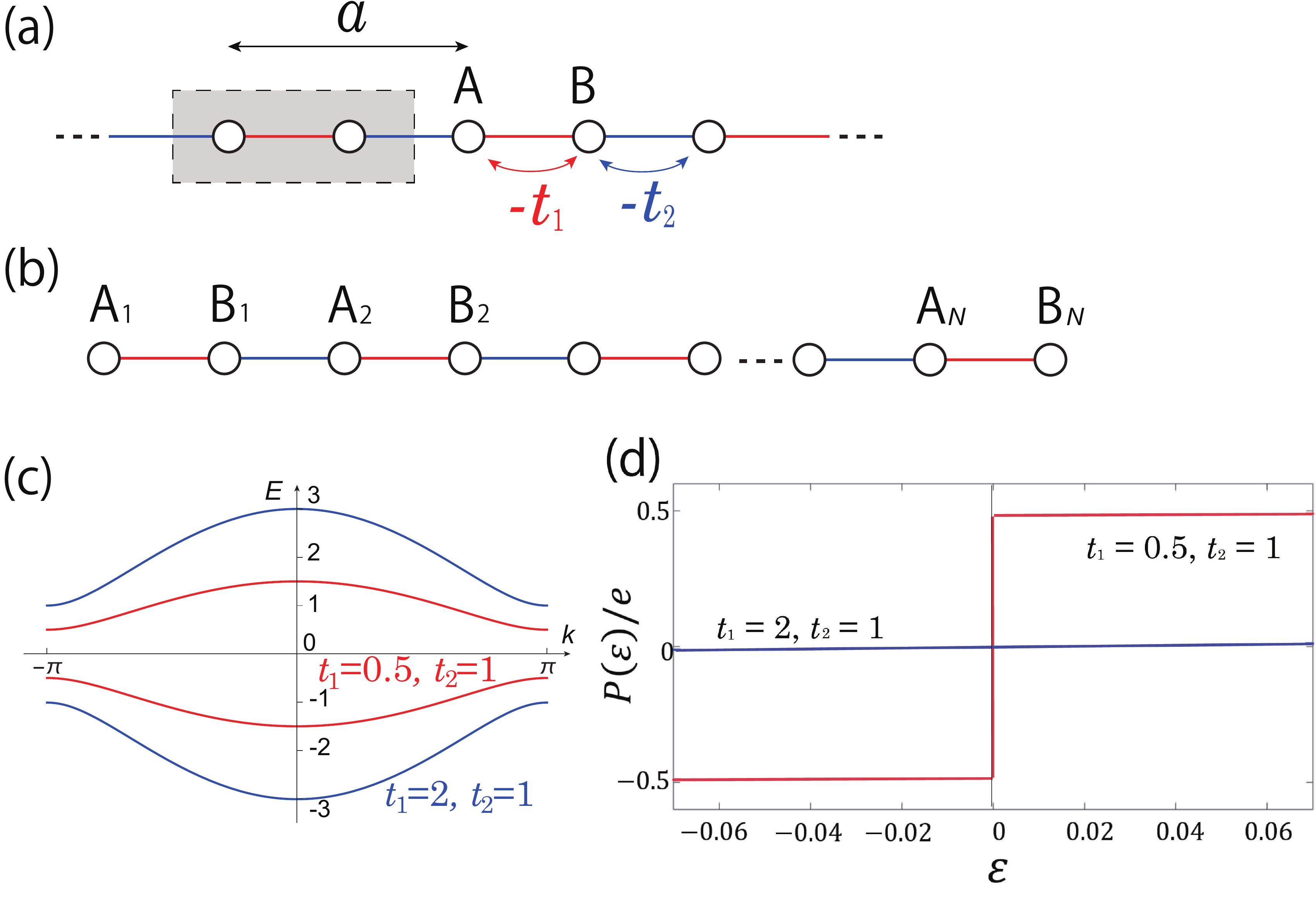}
    \caption{SSH model. (a) Infinite SSH chain. The shaded square represents a unit cell. Hopping amplitudes alternate between the intracell hopping $t_1$ (red lines) and the intercell hopping $t_2$ (blue lines).  (b) Finite open SSH chain. There are $2N$ sites in the chain. (c) 
Band structure for  $t_1=0.5, t_2=1$ (red) and  $t_1=2, t_2=1$ (blue).
(d) 
Dielectric polarization of the SSH model. Here we set $N=20$ and the number of sites is $2N=40$. The hopping amplitudes are $t_1=0.5, t_2=1$ for the red line and  $t_1=2, t_2=1$ for the blue line. In the former case, $P(\varepsilon)$ rapidly changes toward $\pm e/2$ in the vicinity of $\varepsilon=0$.}
 \label{fig:fig1-2}
\end{figure}
%%%%%%%%%%%
%%%
%%%%%%%%%
We first review the definition of the  Zak phase and its connection with electric polarization. For one-dimensional crystalline insulators, the Zak phase is defined as
\begin{align}
\theta=-i\sum_n^{\text{occ.}}\int_0^{2\pi/a}dk\bra{u_n(k)}\dfrac{\partial}{\partial k}\ket{u_n( k)},\label{Zak phase}
\end{align}
where $a$ is the lattice constant, $k$ is the wavenumber, $u_n(k)$ is the periodic part of the Bloch wave function of the $n$th eigenstate, and the summation is over the occupied states. This Zak phase is
defined modulo $2\pi$. It is quantized as 0 or $\pi$ under  the SI symmetry \cite{kariyado2013symmetry}. When $\theta=\pi$, there exist two degenerated edge states in the gap when the system preserves chiral symmetry~\cite{ryu2002topological}. According to the modern theory of polarization \cite{Vanderbilt1,Vanderbilt2}, the electric polarization $P_0$ can be obtained from the Zak phase in the following equation;
\begin{align}
P_0=-e\dfrac{\theta}{2\pi}\ ({\rm mod}\ e).\label{polarization:Zakphase}
\end{align}
This equation indicates that the polarization is $e/2$ (mod $e$) in the system with the $\pi$ Zak phase. 
We note that in one-dimensional systems, the polarization, which is an electric dipole moment per unit length, has the unit of charge. 

   In this section, we consider the SSH model. It is a tight-binding model on a one-dimensional lattice with two sites  $A_{i}$ and $B_{i}$ 
   %per unit cell
   in the $i$th unit cell
   and with different intracell ($-t_1$) and intercell ($-t_2$) hopping amplitudes between $A$ and $B$ sites as shown in Fig.~\ref{fig:fig1-2}(a). We set $t_1$ and $t_2$ to be positive. The distance between the adjacent A site and B site is set to be $a/2$. The Bloch Hamiltonian is given by
\begin{align}
\mathcal H^{\text{bulk}}(k)=\left(\begin{array}{cc}
0&-t_1-t_2e^{-ika}\\
-t_1-t_2e^{ika}&0
\end{array}\right)\ .\label{SSH:BulkHamil}
\end{align} 
This model has chiral symmetry, and therefore the spectrum is symmetric with respect to $E=0$. The energy eigenvalues are given by $E=\pm\sqrt{t_1^2+t_2^2+2t_1t_2\cos ka}$ and the band gap exists when $t_1\ne t_2$ (see Fig. \ref{fig:fig1-2}(c)).
	When $t_1$ and $t_2$ are equal, the gap closes at $k=\pi /a$. 
%	At this time, the size of the unit cell can be halved, so the bands are unfolded at $k= \pi/ a$.
%	

Here we focus on the case of $t_1\ne t_2$, where the system is insulating.
We set the Fermi energy to be $E_f=0$, i.e. the system is half-filled. The Zak phase of the SSH model is calculated by substituting the eigenstates of (\ref{SSH:BulkHamil}) into (\ref{Zak phase}), and we get
\begin{align}
\theta&=\begin{cases}
0\ (t_1>t_2)\\
\pi\ (t_2>t_1)
\end{cases}.\label{ZakPol}
\end{align}
This value of the Zak phase can be calculated from the parity eigenvalues at time-reversal invariant momenta (TRIM), $k=0$ and $k=\pi$:
\begin{align}
e^{i\theta}=\prod_n^{\text{occ.}}\xi_n(k=0)\xi_n(k=\pi),
\label{eq:parity}
\end{align}
where $\xi_n(k)$ is the parity eigenvalue of the $n$th band at $k$, and the product is taken over the occupied bands. 
In the present case of the SSH model, from the inversion operator at the center of the unit cell
$\hat{P}=\left(\begin{array}{cc}0&1\\1&0\end{array}\right)$,
we get 
\begin{align}
\xi(k=0)=1,\ \xi(k=\pi)=\left\{\begin{array}{ll}1&t_1>t_2 \\ -1&t_2>t_1 \end{array}\right.
\end{align}
and we reproduce the result (\ref{ZakPol}). This also means that the bands are inverted at $k=\pi$ when $t_1$ is changed across the gap closing at $t_1=t_2$.

From Eq.~(\ref{polarization:Zakphase}), Eq.~(\ref{ZakPol}) indicates that an electric charge $e/2 \ ({\rm mod}\ e)$ accumulates at both ends when  $t_2>t_1$ meaning that each of the topological edge states are half filled \cite{Vanderbilt1,Vanderbilt2}. 
In order to have such topological edge states due to the $\pi$ Zak phase, the chain should be fully covered by the chosen unit cells. In this way, in order to see bulk-edge correspondence for the Zak phase, the termination of the crystal is related with the choice of the unit 
cell with which the Zak phase is calculated.
We note that appearance of the polarization $P_0$ does not contradict the inversion symmetry because $P_0\equiv e/2\equiv -e/2\ $({\rm mod} $e$). 

%%%
\subsection{Dielectric response}
%%%
We next consider the dielectric polarization in a finite SSH chain (see Fig.~\ref{fig:fig1-2} (b)).  The chain consists of $2N$ sites, and each site is represented as $\sl A_i$ and $\sl  B_i (i = 1,2,\cdots N)$ as shown in the figure. The Hamiltonian of the finite SSH chain is given by
\begin{align}
H_0&=-t_1\sum_{n=1}^{N}a^{\dagger}_{n}b_{n}-t_2\sum_{n=1}^{N-1}a^{\dagger}_{n+1}b_{n}+{\sl h.c.},
\end{align}
where $a^\dagger_i$ and $b^\dagger_i$ are creation operators at the $\sl A_i$ and $\sl B_i$ sites respectively. The Hamiltonian with static electric field $\varepsilon$ is given by
\begin{align}
H&=H_0+H_{\varepsilon},\label{SSH:chain}\\
H_\varepsilon&=\varepsilon ea\sum_{n=1}^N\left[\left(-\dfrac{3}{4}-\dfrac{N}{2}+n\right)a^\dagger_na_n
\right.
\nonumber \\
\ \ \ \ \ \ \ &\left.
+\left(-\dfrac{1}{4}-\dfrac{N}{2}+n\right)b^\dagger_nb_n\right].
\end{align}
The dielectric polarization is given by ;
\begin{align}
P(\varepsilon)=-\dfrac{1}{a(N-1/2)}\sum_{n =1}^{N}\dfrac{\partial E_n}{\partial \varepsilon},\label{SSH:Pol}
\end{align}
where $E_n$ with $n=1,2,\cdots,N$ are the energy eigenvalues of (\ref{SSH:chain}) below $E_f=0$, and $a(N-1/2)$ is the length of the chain. 
Note that unlike the electric polarization $P_0$ determined from the bulk wave function, the dielectric polarization $P(\varepsilon)$ is defined for a chain with a finite length $N$, and is dependent on $N$.
The results for two cases of $t_1=0.5$, $t_2=1$ and $t_1=2$, $t_2=1$ are shown in Fig. \ref{fig:fig1-2}(d), with the band structures given in  
Fig. \ref{fig:fig1-2}(c). 

The result shows that the dielectric polarization takes an anomalously large value of $P(\varepsilon)=\pm e/2$ even for a very weak electric field when $t_1<t_2$. This behavior is dramatically different from that in the other case $t_1>t_2$, where $P(\varepsilon)$ linearly increases with a small slope. 
We note that in the previous work \cite{Combes} the polarization of the Rice-Mele model, which is 
the SSH model with a staggered potential added, is calculated with the 
same formalism, and its relation to the Zak phase is discussed. Nonetheless, an 
abrupt change of the polarization as a function of the electric field in the SSH model, 
which we find in this paper, is not studied in Ref.~\onlinecite{Combes}.
Differences between the present work and Ref.~\onlinecite{Combes} will be discussed in 
Sec.~\ref{sec:bulk-surface}.

\begin{figure}[tbp]
    \centering
    \includegraphics[width=86mm]{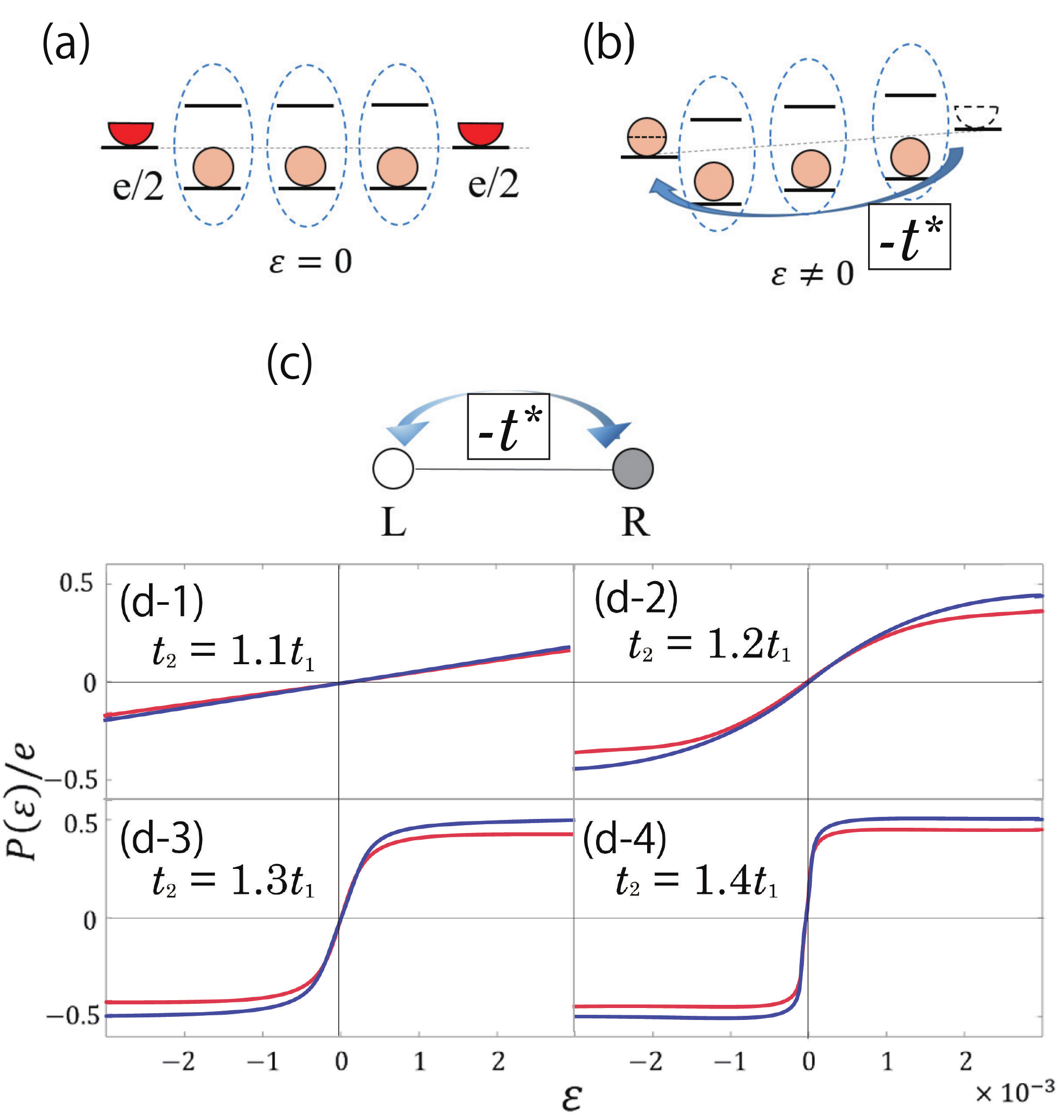}
    \caption{Dielectric response of the SSH model (Model I). (a)(b) Schematic picture of the anomalous dielectric polarization for $t_2\gg t_1$. In such a case, the levels in each dimer formed by the strong bonds are split by the hopping $t_2$, and there appears an isolated state at each end of the chain. The orange circles represent electrons and red semicircle represents a half of an electron. In (a), the electric field $\varepsilon$ is zero, and the states at each end is half-filled. In (b), the electric field is applied, and there appears a difference between the levels at the two ends, and a half of an electron charge is carried from one end to the other. (c) Schematic picture of the effective two-site model. (d-1)-(d-4) Dielectric polarization for the SSH model (red lines) and effective two-site model (blue lines). The hopping amplitudes are taken as $t_1=1$, (d-1) $t_2=1.2$, (d-2) $t_2=1.3$,  (d-3) $t_2=1.4$, and (d-4) $t_2=1.5$. }
 \label{fig:fig3-4}
\end{figure}

 The $\pi$ Zak phase indicates presence of topological edge states, and we attribute this anomalous dielectric response to these edge states. Because of the chiral symmetry, the edge state at each end of the chain is at the zero energy, when hybridization between the edge states at the two ends is neglected. These edge states are half filled. Thus, at an either edge, there appears an edge charge equal to $e/2$ as shown in Fig. \ref{fig:fig3-4}(a). We attribute the anomalous dielectric response for the insulating phase with the $\pi$ Zak phase to these charges at the edge states. When a small electric field $\varepsilon$ is applied, the two edge states will have different energies as shown in Fig.~\ref{fig:fig3-4}(b), giving rise to an immediate transfer of charges from one end to the other. This concept is verified using an effective two-site model (Fig. \ref{fig:fig3-4}(c)). This model consists of two sites, each of which corresponds to one of the two edge sites of the SSH model. The parameter $t^\ast$ represents an effective hopping amplitude between the two sites, L and R, representing the left and right ends of the chain, respectively. Let $\ket{R}=(1,0)^T \left [\ket{L}=(0,1)^T ]\right)$ denote the state where the electron is at the R [L] site. Then, the effective Hamiltonian with the electric field takes the form
\begin{align}
H^{\ast}=\left(\begin{array}{cc}
\varepsilon ea(2N-1)/4& -t^\ast\\
-t^\ast &-\varepsilon ea(2N-1)/4
\end{array}\right)\ .\label{SSH:efH}
\end{align}
We put one electron into the model. The effective dielectric polarization is given by
\begin{align}
P^\ast(\varepsilon)&=\frac{e}{2}\left[\braket{R|-}\braket{-|R}-\braket{L|-}\braket{-|L}\right]\notag\\
&=-\text{sgn}(\varepsilon)\dfrac{e}{2\sqrt{(\frac{2t^\ast}{(N-1/2)ae\varepsilon})^2+1}}\ \label{SSH:EffPol},
\end{align}
where $\ket{-}$ is the eigenstate of Eq.~(\ref{SSH:efH}) with the negative eigenvalue. The hopping parameter $t^\ast$ is determined by fitting Eq.~(\ref{SSH:EffPol}) for $P^\ast(\varepsilon)$ to our numerical result $P(\varepsilon)$ within the region of a weak electric field. The comparisons between (\ref{SSH:Pol}) and (\ref{SSH:efH}) are shown in Fig. \ref{fig:fig3-4}(d). We can see the two graphs fit well even for a strong electric field and evolve in the same way as $t_2/t_1$ changes. This result confirms that the anomalous dielectric response is caused by the topological edge states. 
We also note that this response around $\varepsilon\sim 0$ becomes sharper for a longer chain, because in Eq.~(\ref{SSH:EffPol}) 
$t^*$ becomes smaller and $N-1/2$ becomes larger for a longer chain.

%%%FIG3%%%%

%%%%%%%%%%%%%%%%%%%%%%%%%%%%%%%%%%%%%%%%%
\section{three-dimensional model (Model II)}\label{3D}
%%%%%%%%%%%%%%%%%%%%%%%%%%%%%%%%%%%%%%%%%
The discussion in the previous section on a one-dimensional model can be extended to higher-dimesions. 
In this section, 
we show that a jump of the dielectric response also appears in 
a simple three-dimensional tight-binding model on the diamond lattice with anisotropic hopping amplitudes, which we 
call Model II.
In addition, we show other examples showing the similar effect, two two-dimensional models and one three-dimensional model in Appendices A and C, respectively. 
In three-dimensional systems, the polarization, which is an electric dipole moment per unit volume, has the unit of charge divided by an area. 

\begin{figure}[tbp]
    \centering
    \includegraphics[width=86mm]{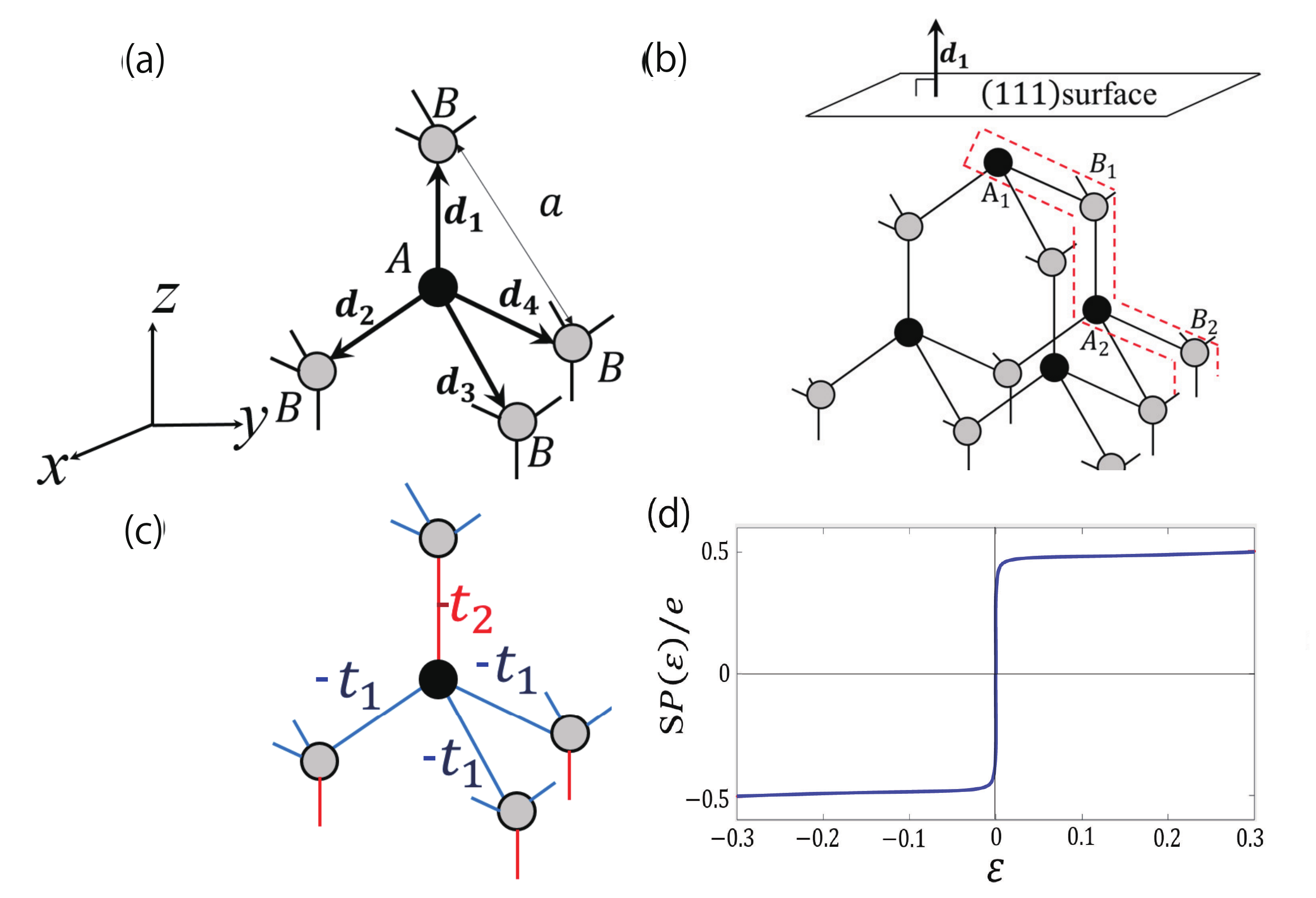}
    \caption{Model II on the diamond lattice. (a) Diamond lattice. The lattice constant is $a$ and the nearest bond vectors from site A to B are $\bm d_{i=1,2,3,4}$. (b) Slab of the model on the diamond lattice with a (111) surface. The red dotted line shows the unit cell. (c) Schematic figure of the model with anisotropic hopping amplitude (Model II). The hopping amplitude $-t_2$ on the bond along (111) direction is different from that along other directions, $-t_1$. (e) Dielectric polarization of a slab of the Model IV on a diamond lattice with (111) surfaces. The polarization $P(\varepsilon)$ rapidly changes toward $\pm e/(2S)$ in the vicinity of $\varepsilon=0$. $S$ is the area of the surface unit cell. Here we set $N=10$ and the number of sites in the unit cell is $20$. The hopping amplitudes are $(t_1,t_2)=(1,4)$.}
    \label{fig:ModelII}
\end{figure}

The diamond lattice and its slab with (111) surfaces are shown in Fig.~\ref{fig:ModelII} (a) and (b). The lattice consists of two sublattices ${\rm A}$ and ${\rm B}$. Let $a$ denote the distance between two neighboring ${\rm A}$ sites. The nearest bond vectors $\bm d_{i=1,2,3,4}$ from an A site to the adjacent B sites are given by
\begin{align} 
\bm d_1&=\frac{\sqrt{6}a}{4}(0,0,1), \bm d_2=\frac{\sqrt{6}a}{4}(-\frac{2\sqrt{2}}{3},0,-\frac{1}{3}),\notag \\
\bm d_3&=\frac{\sqrt{6}a}{4}(\frac{\sqrt{2}}{3},\frac{{\sqrt{6}}}{3},-\frac{1}{3}),\ \bm d_4=\frac{\sqrt{6}a}{4}(\frac{\sqrt{2}}{3},-\frac{{\sqrt{6}}}{3},-\frac{1}{3}),\notag \\
\end{align}
as shown in Fig. \ref{fig:ModelII}. 

%%%%%%%%%%%

%%%
\subsection{Zak phase and dielectric polarization for three-dimensional systems}
First, we review the Zak phase and its connection with electric polarization in three-dimensional systems. When we take the integration path along the  reciprocal lattice vector $\bm G_\perp$, the Zak phase is defined as
\begin{align}
\theta(\bm k_\parallel)=-i\sum_n^{\text{occ.}}\int_0^{|\bm G_\perp|}dk_\perp\bra{u_n(\bm k)}\dfrac{\partial}{\partial k_\perp}\ket{u_n(\bm k)},\label{3DZak phase}
\end{align}
where $\bm k =k_\perp \bm n +\bm k_\parallel$, $\bm n =\bm{G}_{\perp}/|\bm{G}_{\perp}|$ and ${\bm k}_{\parallel}\perp{\bm G}_{\perp}$. The Zak phase is related with the polarization for the systems with surface perpendicular to $\bm G_\perp$. This phase is quantized as 0 or $\pi$ modulo $2\pi$ under both the TR and SI symmetries \cite{kariyado2013symmetry}. In the region of the surface Brillouin zone where the Zak phase is $\pi$, there exist degenerated surface states at zero energy when the chiral symmetry is present \cite{ryu2002topological}. The electric polarization $P_0$ along ${\bm n}$ can be obtained from the Zak phase as shown in the following equation;
\begin{align}
P_0=-e
\int_{\bm k_\parallel\in \text{2DBZ}}
\dfrac{d^2 \bm k_\parallel}{(2\pi)^3}\theta(\bm k_\parallel)\ ({\rm mod}\ e/S),
\end{align}
where $S$ is the area of surface unit cell, and the integral is over the two-dimensional 
Brillouin zone for $\bm k_\parallel$.

\subsection{Model II: Anisotropic tight-binding model on the diamond lattice}\label{DSAA}

We consider a model as shown in Fig.~\ref{fig:ModelII} (c). The system has anisotropic hopping amplitudes; $-t_2$ is the hopping amplitude along the bonds along the (111) direction, and $-t_1$ is that along the other directions. We set $t_1$ and $t_2$ positive. The Hamiltonian is given by
\begin{align}
H_0=-\sum_{\langle ij \rangle}t_{ij}c^{\dagger}_ic_j\ ,\label{HamilOfNonEqSla}
\end{align}
where $t_{ij}$ is the hopping amplitude from site $i$ to $j$. First, we consider a bulk system to calculate the Zak phase. The bulk Bloch Hamiltonian is given by
\begin{align}
H^{\text{bulk}}(\bm k)&=\left(\begin{array}{cc}
0&R(\bm k)\\
R^\ast(\bm k)&0
\end{array}\right),\\
R(\bm k)&= -t_1e^{-i{\bm k}\cdot{\bm d_1}}-t_2(e^{-{\bm k}\cdot {\bm d_2}}+e^{-i{\bm k}\cdot {\bm d_3}}+e^{-{\bm k}\cdot {\bm d_4}}),
\end{align}
where $\bm k=(k_x,k_y,k_z)$. The band gap is open when $t_2>3t_1$ \cite{takahashi2013completely}, and it is around $E=0$. We set the Fermi energy to be $E_f=0$. Now we calculate the Zak phase $\theta(\bm k_\parallel)$ along ${\bm G}_{\perp}=(0,0,\sqrt{6}\pi/a)$ for $\bm k_\parallel=(k_x,k_y)$. Because of the time-reversal and inversion symmetries, the Zak phase is quantized as $0$ or $\pi$ modulo $2\pi$. As an example, we calculate $\theta_z(\bm k_\parallel=0)$ from the products of the parity eigenstates at the $\Gamma$ ($\bm{k}=0$) and the $L$ ($\bm{k}=(0,0,\sqrt{6}\pi/(2a)$) points in the following way \cite{hughes2011inversion}
\begin{align}
e^{i\theta_z(\bm k_\parallel=0)}=\prod_n^{\text{occ.}}\xi_n(\Gamma)\xi_n(L)=-1 ,
\end{align}
because these two points are time-reversal invariant momenta on the $\bm k_\parallel=0$ line.
In this calculation, the unit cell is taken to be a pair of the A and B sites displaced by ${\bm d_2}$, as determined from the crystal termination in this case, and the inversion center is in the middle of the unit cell.
Therefore this equation indicates $\theta_z(\bm k_\parallel)\equiv\pi$ for any $\bm k_\parallel$ because the Zak phase is constant in the two-dimensional Brillouin zone for ${\bm k}_{\parallel}$ when the band gap is open.

The Hamiltonian of the slab with the (111) surfaces (Figs.~\ref{fig:ModelII} (b) and (c)) is given by
\begin{align}
H_0(\bm k_\parallel)&=-\left[t_2\sum_{n=1}^{N-1}e^{i\tilde {\bm d_1}\cdot{ \bm k}_\parallel}a^\dagger_{\bm k_\parallel,n+1}b_{\bm k_\parallel,n} \right. \notag \\
&\ \ \  +\left. t_1\sum_{n=1}^N\left(e^{i\tilde {\bm d_2}\cdot {\bm k_\parallel}}+ e^{i\tilde {\bm d_3}\cdot{\bm k_\parallel}}+ e^{i\tilde {\bm d_4}\cdot{\bm k_\parallel}}\right)a^\dagger_{\bm k_\parallel,n}b_{\bm k_\parallel,n} \right]\notag\\
&\ \ \  +h.c.,
\end{align}
where $a_{\bm k_\parallel,n}^\dagger(b_{\bm k_\parallel,n}^\dagger)$ is the creation operator of an electron  with Bloch wave vector $\bm k_\parallel$ at the $n$th A(B) site in the unit cell and $\tilde{\bm d_{i}}=(d_{ix},d_{iy})$. When we apply an electric field of strength $\varepsilon$ along the $z$ axis, the Hamiltonian is given by
\begin{align}
H(\bm k_\parallel)&=H_0(\bm k_\parallel)+H_\varepsilon(\bm k_\parallel),\label{HamNonEqSlabElec}\\
H_\varepsilon(\bm k_\parallel)&=\dfrac{\sqrt{6}}{4}ae\varepsilon\sum_{n=1}^N\left[ \left(\dfrac 4 3 n -\dfrac 2 3 N -\dfrac 5 6 \right)a^\dagger_{\bm k_\parallel,n}a_{\bm k_\parallel,n} \right.\notag \\
&\ \ \ \ \ \ \ \left.+ \left(\dfrac 4 3 n -\dfrac 2 3 N -\dfrac 1 2\right)b^\dagger_{\bm k_\parallel,n}b_{\bm k_\parallel,n}\right],
\end{align}
 The dielectric polarization is given by 
\begin{align}
P(\varepsilon)=-\dfrac{1}{a(4N/3-1)}\sum_{n=1}^N\int_{\bm k_\parallel\in \text{2DBZ}}\dfrac{d^2\bm k_\parallel}{(2\pi)^2}\dfrac{\partial E_n(\bm k_\parallel)}{\partial \varepsilon},
\end{align}
where the integral is taken over the two-dimensional Brillouin for ${\bm k}_{\parallel}$, $E_n(\bm k_\parallel)$ is the $n$th lowest eigenvalue of (\ref{HamNonEqSlabElec}), and $a(4N/3-1)$ is the thickness of the system. The calculation results are shown in Fig.~\ref{fig:ModelII}(e). The dielectric polarization takes a value $P(\varepsilon)\sim\pm e/2S$ even for a very weak electric field, where $S$ is an area of the surface unit cell. This indicates that a half of an electron per surface unit cell is accumulated on each surface  whenever $t_2>3t_1$, and the response of this surface charge gives rise to the anomalous behavior of the polarization. Thus, the anomalous dielectric response occurs in this system when it is insulating.

%%%
\section{Remarks on various effects on the anomalous dielectric responses}
\label{sec:remark}
We have discussed several models in one, and three dimensions with the $\pi$ Zak phase, and demonstrated that there is an abrupt jump in the dielectric response. 
Because they are idealized models, we make several remarks in order to see how the anomalous dielectric responses in general systems including in real materials. 

\subsection{Effect of surface-state dispersion}
\label{sec:dispersion}
Here, we discuss  flatness of surface-state dispersions and its relation to anomalous dielectric response. 
In the models in Secs. II and III and in Appendices A and C, the edge/surface states are exactly at the zero energy and are dispersionless, which is due to chiral symmetry and particle-hole symmetry in these models. 
Under the chiral symmetry the energy band becomes symmetric with respect to $E\leftrightarrow -E$, which means that the
energy eigenvalues satisfy $E_{n,\mathbf{k}}=-E_{m,\mathbf{k}}$. 
Under the particle-hole symmetry the energy eigenvalues satisfy $E_{n,\mathbf{k}}=-E_{m,\mathbf{-k}}$. 
Because of the inversion symmetry of the sysem, both the chiral symmetry and the particle-hole symmetry 
are preserved here, which makes the topological boundary states to be exactly at the zero energy. 
Therefore, as we discussed so far, their response under a small electric field becomes anomalous, since even for a small electric field, 
the energies of all the states on one surface become positive, while those on the other surface become negative, and all the surface charges 
are transferred at the same time. 
In contrast, general systems do not have chiral or particle-hole symmetry, and the edge/surface states have dispersions. 
This edge-/surface-state dispersion will make the dielectric response less abrupt, because upon the application of the electric field, not all the charges in these edge/surface states are
carried from one end to the other at the same time. Thus depending on the dispersion of the edge/surface states, the response becomes less anomalous. 
We show several examples for dispersions of surface states in real materials with the $\pi$ Zak phase in the next section.

Thus far we have seen that the polarization jumps to $e/(2S)$ by a weak electric field when the surface-state dispersion is flat.
In this discussion we do not take into account electron correlation. If we take electron-electron interaction into account, 
the dielectric response may change. When a weak electric field is applied, the electrons at the surface may segregate and form a superstructure,
so that the correlation energy may overcome the potential difference of the surface states at the both ends of the slab. Such kinds of 
formation of charge segregation will make the dielectric response less abrupt. 
We do not argue this phenomena any further here, because it should be sensitive to various details of the system
such as the form and the strength of the electron-electron interaction, dispersion of the surface states, and so on. 

\subsection{Effect of the depolarization field}
 So far we have ignored an influence of the depolarization field. When an electric field is applied and a polarization is induced, the induced polarization generates a depolarization field. This depolarization field depends on the geometry of the system. Particularly when the shape of the system is an ellipsoid or a large thin slab, the influence can be easily calculated because the depolarization field is uniform, represented by  the depolarization factor. 

\begin{figure}[tbp]
 \centering
  \includegraphics[width=80mm]{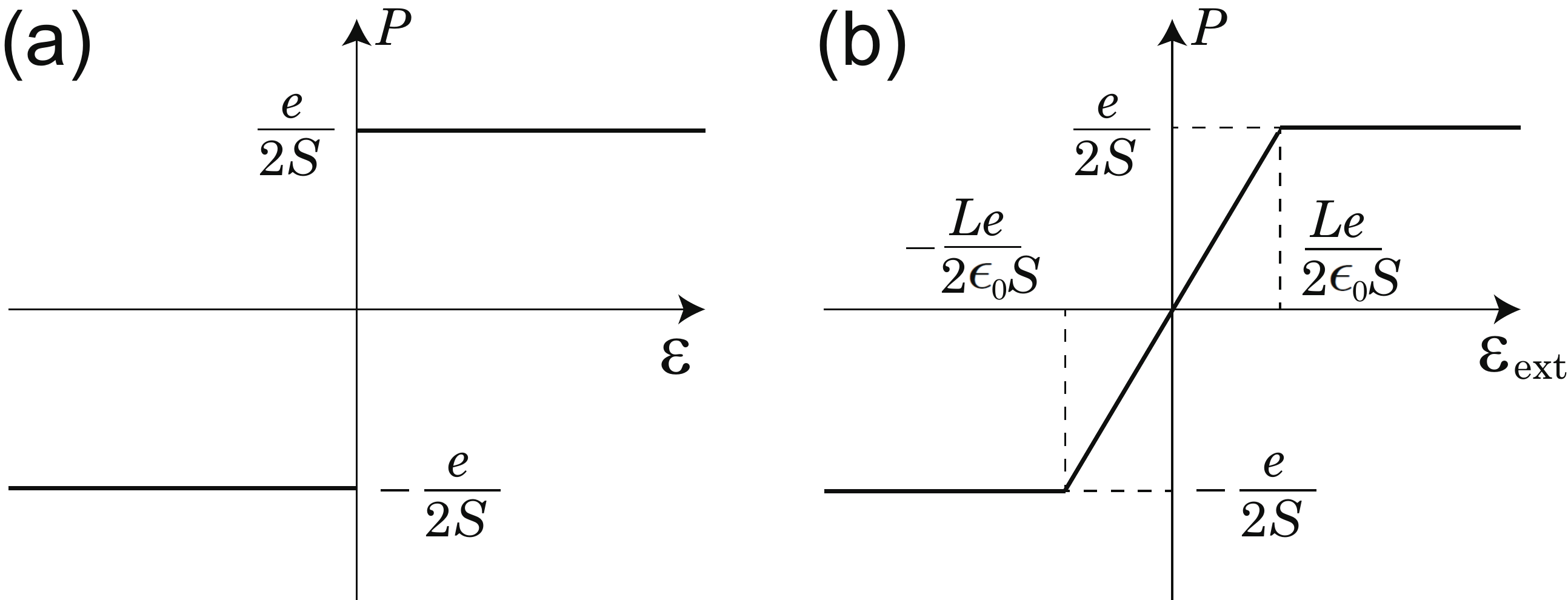}
\caption{Effect of depolarization field for an ideal case, where the Zak phase equal to $\pi$ and the dispersion of the topological surface states is flat. (a) Polarization $P$ versus the electric field $\varepsilon$. 
(b) Polarization $P$ versus the external electric field $\varepsilon_{\rm ext}$. 
}
% \caption{Dielectric polarization calculated as a response to the external electric field. Here we set $N=10$ and the number of sites in the unit cell is $20$. The Slater-Koster parameters are $(V_{ss},V_{sp},V_{pp\sigma},V_{pp\pi})$ =  $(-4.43,3.79,5.66,-1.83)$. The polarization $P(\varepsilon)$ increases to $\pm e/2S$ and the slope on the graph is $\epsilon_0$. We put the depolarization factor to be $L=1$, corresponding to %a thin slab perpendicular to the electric field. }
 \label{fig:depol}
\end{figure}

The electric field $\varepsilon$ used in the discussion so far represents the total electric field felt by the electrons.  It is a sum of the external electric field $\varepsilon_{\text{ext}}$ and the depolarization field $\varepsilon_{\text{dep}}$. The depolarization field $\varepsilon_{\text{dep}}$ is expressed as $\varepsilon_{\text{dep}}=-LP/\epsilon_0$, where $P$ is the dielectric polarization, $L$ ($0\leq L\leq1$) is the depolarization factor, and $\epsilon_0$ is the permittivity of vacuum. Therefore, the total electric field $\varepsilon$ takes the form;
\begin{align}
\varepsilon=\varepsilon_{\text{ext}}+\varepsilon_{\text{dep}}=\varepsilon_{\text {ext}}-LP(\varepsilon)/\epsilon_0, 
\label{eq:dep}
\end{align}
Here, the dependence of $P(\varepsilon)$ on $\varepsilon$ has been discussed so far in this present paper. 
%Thus, $\varepsilon_{\text ext}=\varepsilon+LP(\varepsilon)/\epsilon_0$ gives the dielectric response to the external field.
Now, we focus on the case with the Zak phase equal to $\pi$ for any value of $\bm{k}_{\|}$, and 
for simplicity we assume that 
the topological surface states have flat dispersion. Then, the response of the polarization $P$ to the electric field $\varepsilon$ is 
given by 
\begin{align}
P=\frac{e}{2S}\text{sgn}(\varepsilon),
\end{align}
where $S$ is an area of the surface unit cell, as shown in Fig.~\ref{fig:depol}(a). From this equation, and Eq.~(\ref{eq:dep}), the dielectric response $P$ versus the external electric field $\varepsilon_{\text{ext}}$ 
is shown in Fig.~\ref{fig:depol}(b). Thus, the slope around $\varepsilon_{\text{ext}}\sim 0$ is $\epsilon_0/L$. 

The present system corresponds to a dielectric with the dielectric constant 
$\epsilon=\infty$, because a nonzero polarization 
arises in response to an infinitesimal electric field. As a comparison, let us consider a
uniform dielectric with a dielectric constant $\epsilon$. Then by including the effect of the 
depolarization field, with the depolarization factor $L$, its dielectric response is given by 
\begin{align}
P=(\epsilon-\epsilon_0)\varepsilon,\ \varepsilon=\varepsilon_{\text{ext}}-\frac{LP}{\epsilon_0}
\end{align}
Thus the ratio of the polarization to the external electric field $\varepsilon_{\text{ext}}$, 
is given by $P/\varepsilon_{\text{ext}}=\frac{\epsilon_0(\epsilon-\epsilon_0)}{L(\epsilon-\epsilon_0)+\epsilon_0}$. By comparison with our result in 
Fig.~\ref{fig:depol}(b), the present system indeed corresponds to the dielectric with $\epsilon=\infty$. 
%
 %It is plotted in Fig. \ref{fig:PolInt} with $L=1$, corresponding to a thin slab perpendicular to the electric field. The result shows that the dielectric polarization linearly increases with a slope $\epsilon_0$ as a response to the external electric field.

%%%%%%%%%%%%%%%%%
%%%%%%%%%%%%%%

\subsection{Dynamical Dielectric Response in an insulator with the $\pi$ Zak phase}
\begin{figure}[tbp]
 \centering
  \includegraphics[width=70mm]{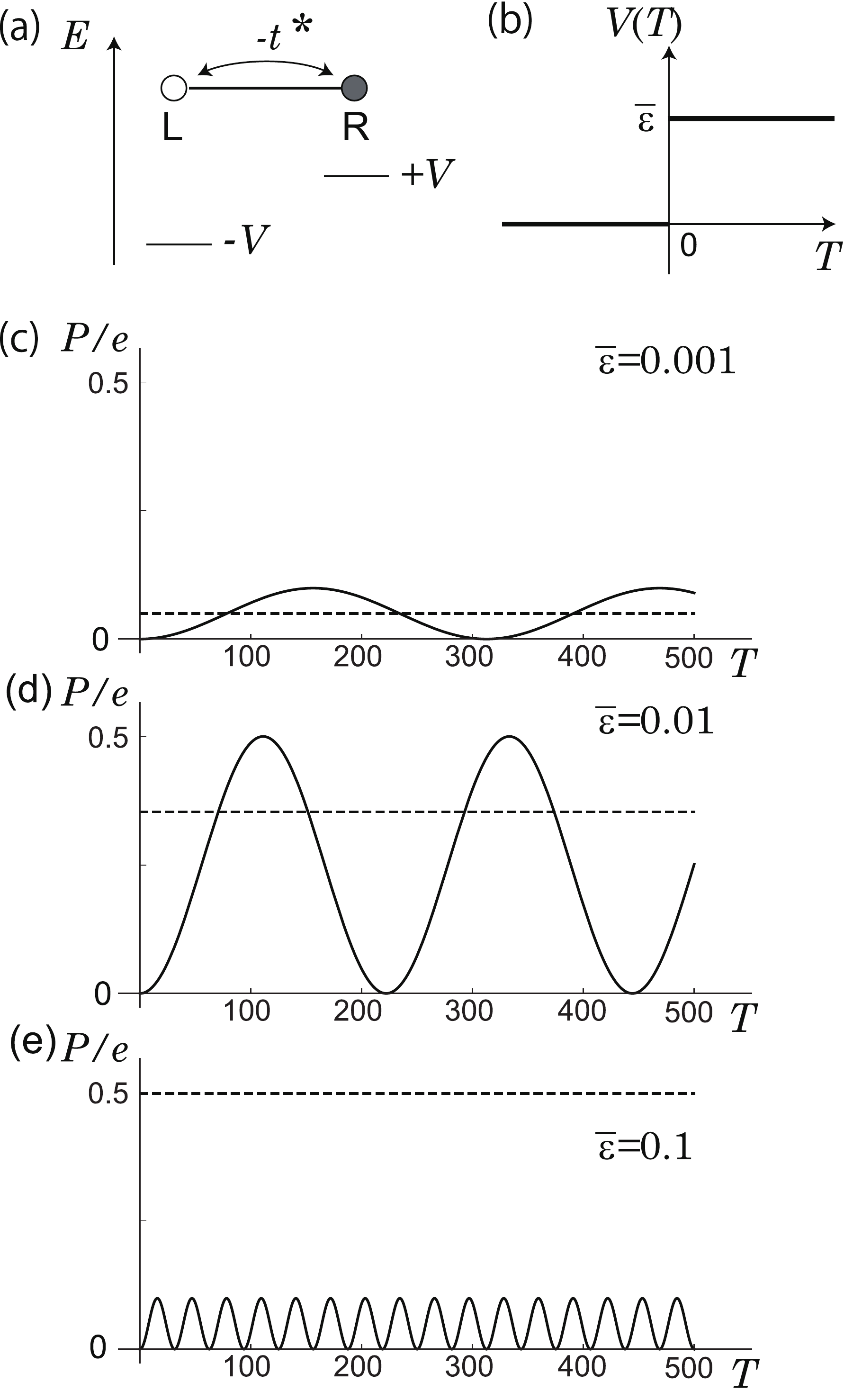}
 \caption{Dynamics of polarization under the electric field.
 (a) Schematic figure of the setup. (b) External electric field used in the calculation, which is turned on at $T=0$. (c)-(e) Time dependence of the dynamical polarization wih $t^*=10^{-2}$ and $\bar{\varepsilon}=$ (c) $10^{-3}$, (d) $10^{-2}$, and (e) $10^{-1}$. 
 The solid lines show the dynamic response $P^*_{dyn.}(T)$ and the broken lines show the static response $P^*_{sta.}$.}
 \label{fig:dynamics}
\end{figure}

So far we have discussed how anomalous dielectric response
appears in a static case. We here discuss dynamical response; we
suppose there is no electric field when the time $T$ is in 
$T<0$, and then we apply a
constant electric field when $T\geq 0$. 
To study dynamical response we study the following model, 
representing the two states at the two ends of the one-dimensional 
system similar to Eq.~(\ref{SSH:efH}):
\begin{equation}
H=\left(\begin{array}{cc}
V(T)&-t^{*}\\
-t^*&-V(T)
\end{array}
\right).
\end{equation}
Here, we set $\ket{R}=(1,0)^T \left [\ket{L}=(0,1)^T ]\right)$ to denote the state at the right (left) end,
similarly to Eq.~(\ref{SSH:efH}), as shown in Fig.~\ref{fig:dynamics}(a).
In addition, $-t^*$ ($t^*>0$) is the hopping between the two states at the two ends, and $\pm V(T)$ are the energies of these two states, due to the external electric field. Here $t^*$ is an exponentially decaying function of the chain length. 
When $T<0$, the external electric field is zero and $V(T)=0$, and when $T>0$, the external 
electric field is applied and $V(T)=\bar{\varepsilon}$, where $\varepsilon$ is a 
constant proportional to the electric field (Fig.~\ref{fig:dynamics}(b)). 
We set the Fermi energy $E_F=0$ so that only one state is 
occupied. At the time $T=0$, the occupied state is expressed as
\begin{equation}
\psi(T=0)=\frac{1}{\sqrt{2}}\left(\begin{array}{c}1\\1\end{array}\right).
\end{equation}
The evolution of this state is described by the time-dependent
Schr\"odinger equation
\begin{equation}
i\hbar\frac{\partial}{\partial T}\psi(T)=H\psi(T), 
\end{equation}
Through a straightforward calculation, we obtain
\begin{equation}
\psi(T)=\frac{1}{2\sqrt{2}A}\left(\begin{array}{c}
e^{-iAT/\hbar}(A-t^*+\bar{\varepsilon})+e^{iAT/\hbar}(A+t^*-\bar{\varepsilon})
\\
e^{-iAT/\hbar}(A-t^*-\bar{\varepsilon})+e^{iAT/\hbar}(A+t^*+\bar{\varepsilon})
\end{array}\right),
\end{equation}
for $T>0$, where $A=\sqrt{\bar{\varepsilon}^2+t^{*2}}$.
Thus the probablity for the electron to reside at the left (right) end of the chain is
calculated as 
\begin{equation}
n_{R(L)}(T)=\frac{1}{2}\left(1\mp \frac{2t^{*}\bar{\varepsilon}}{A^2}\sin^2\frac{2AT}{\hbar}\right)
\end{equation}
Therefore, the dynamical polarization $P^*_{dyn.}$ is given by
\begin{equation}
P^*_{dyn.}(T)=-\frac{e}{2}(n_{R}-n_{L})=\frac{e}{2}\frac{2t^{*}\bar{\varepsilon}}{A^2}\sin^2\frac{2AT}{\hbar}
\end{equation}

On the other hand, the static polarization is also calculated in response
to the static electric field, similarly to Sec.~IIB. By putting $V\equiv \bar{\varepsilon}$, the occupied state
is given by
\begin{equation}
\Psi=\frac{1}{\sqrt{2A(A+\bar{\epsilon})}}\left(\begin{array}{c}
t^{*}\\A+\bar{\epsilon}
\end{array}\right),
\end{equation}
Thus the probablity for the electron to reside at the left (right) end of the chain is
calculated as 
\begin{equation}
n_{R(L)}=\frac{1}{2}\left(1\mp \frac{\bar{\varepsilon}}{A}\right)
\end{equation}
Therefore, the static response of the polarization $P^*_{sta.}$ is given by
\begin{equation}
P^*_{sta.}=-\frac{e}{2}(n_{R}-n_{L})=\frac{e}{2}\frac{\bar{\varepsilon}}{A}
\end{equation}
The dynamical and static polarizations are shown in Fig.~\ref{fig:dynamics}, 
for $t^*=10^{-2}$ and $\bar{\varepsilon}=10^{-3}, 10^{-2}, 10^{-1}$. 

The dynamical polarization $P^*_{dyn.}$ undergoes Rabi oscillations. 
When $\bar{\varepsilon}=10^{-3}\ll t^{*}$, the polarization oscillates with 
its time average almost equal to the static value $\langle P_{dyn.}^*(T)\rangle\sim P^*_{sta.}\sim\frac{e}{2}\frac{\bar{\varepsilon}}{t^*}$
On the other hand, when $\bar{\varepsilon}=10^{-1}\gg t^{*}$, the dynamical polarization with the Rabi oscillation 
is much smaller than the static value. This Rabi oscillation
appears due to the interference between two eigenstates. 
Therefore, when we include dissipation to other degrees of freedom such as phonons, 
the state showing the Rabi oscillation will gradually relax to the lowest-energy state of the static Hamiltonian, and this oscillating polarization $P^*_{dyn.}$ will gradually approach 
the value of the static one $P^*_{sta.}$. 
Nonetheless, in the present context, the overlap between the two states at the two ends of the chain is 
small, and the dissipation from the higher-energy state to the ground state is very slow. Thus 
it may take a long time to approach the static value of the polarization, which shows anomalous behaviors proposed in the previous Sections.

%%%%%
%%%%%%%%%%

\section{Proposals for real materials}
\label{sec:material}
In this section, we propose some materials whose Zak phase has a quantized value of $\pi$ along some direction. In such cases, there exist topological in-gap states, similar to the in-gap states
at the two ends of the SSH model within the topological regime. 
To search for candidate materials, we point out that the $\pi$ Zak phase typically means existence of dangling bonds. One can see it from the simple example of the SSH model. The $\pi$ Zak phase 
and resulting in-gap states appear when the bond within each dimer is cut at the end of the system. This is nothing but a dangling bond. 
It is also similar in two and three dimensions. The $\pi$ value of the Zak phase is realized when the Wannier orbitals are ``cut into halves'' at the surfaces. In covalent crystals, where the Wannier orbitals are in the middle of the bonds, this typically leads to dangling bonds. On the other hand, if the Wannier orbitals are at the atomic sites, one cannot cut the crystal at the center of the Wannier orbitals. Thus, realization of a surface supporting the in-gap surface states with $\pi$ Zak phase is not straightforward from the viewpoint of stability of materials.

Here we propose three classes of materials, which circumvent this obstacle against realizing the in-gap surface states due to the $\pi$ Zak phase. 
One is a cubic semiconductor with the (111) surface, such as diamond and silicon. In this case, the surface has in-gap surface states due to the $\pi$ Zak phase, if the surface reconstruction is
absent. In reality, dangling bonds exist at the surface, leading to surface reconstructions. Nevertheless, as we show later, when the surface reconstruction leads to the formation of surface superstructure with enlarged surface unit cell by $M$ times, where $M$ is an integer, the Zak phase becomes $M\pi$ modulo $2\pi$. 
Therefore, when $M$ is odd, the Zak phase remains $\pi$ and the topological in-gap states survive. 

Another example is a carbon nanotube.
In this case, the covalent bond of the $sp_2$ hybrid orbital gives rise to the $\pi$ Zak phase.
In nanotubes, existence of a gapless state and symmetry of the system change depending on the chirality $(m, n)$.
Here, we focus on a zigzag nanotube with chirality $(m,0)$ with $m$ being an odd number other than a multiple of 3, so that it is insulating and the Zak phase is $\pi$.

The other material is Sc$_2$C, one of the topological electrides \cite{hirayama2018electrides}. It is a layered material, and the Wannier orbitals reside between the atomic layers; therefore, 
one can create the surface by cutting the system between the atomic layers, leading to the $\pi$ Zak phase.

\subsection{Diamond and silicon}\label{Ab initio calculations for diamonds}
In this section, we show that diamond and silicon  have the $\pi$ Zak phase
along the [111] direction over the entire two-dimensional Brillouin zone along the (111) surface, similarly to Sec.~\ref{3D}. 
In fact, this was first pointed out in Ref.~\cite{Vanderbilt2}. 
In Ref.~\cite{Vanderbilt2}, various cubic semiconductors are studied, and it was found that their (111) and 
($\bar{1}\bar{1}\bar{1}$) surfaces have one half of an electron per 1$\times$ 1 surface unit cell (excluding the spin degree of freedom), 
corresponding to the $\pi$ Zak phase. 
%The surface reconstruction, which we have not considered in the discussion so far, occurs in these materials, and we show that our result is applicable also in this case.

\begin{figure}[tbp]
 \centering
  \includegraphics[width=86mm]{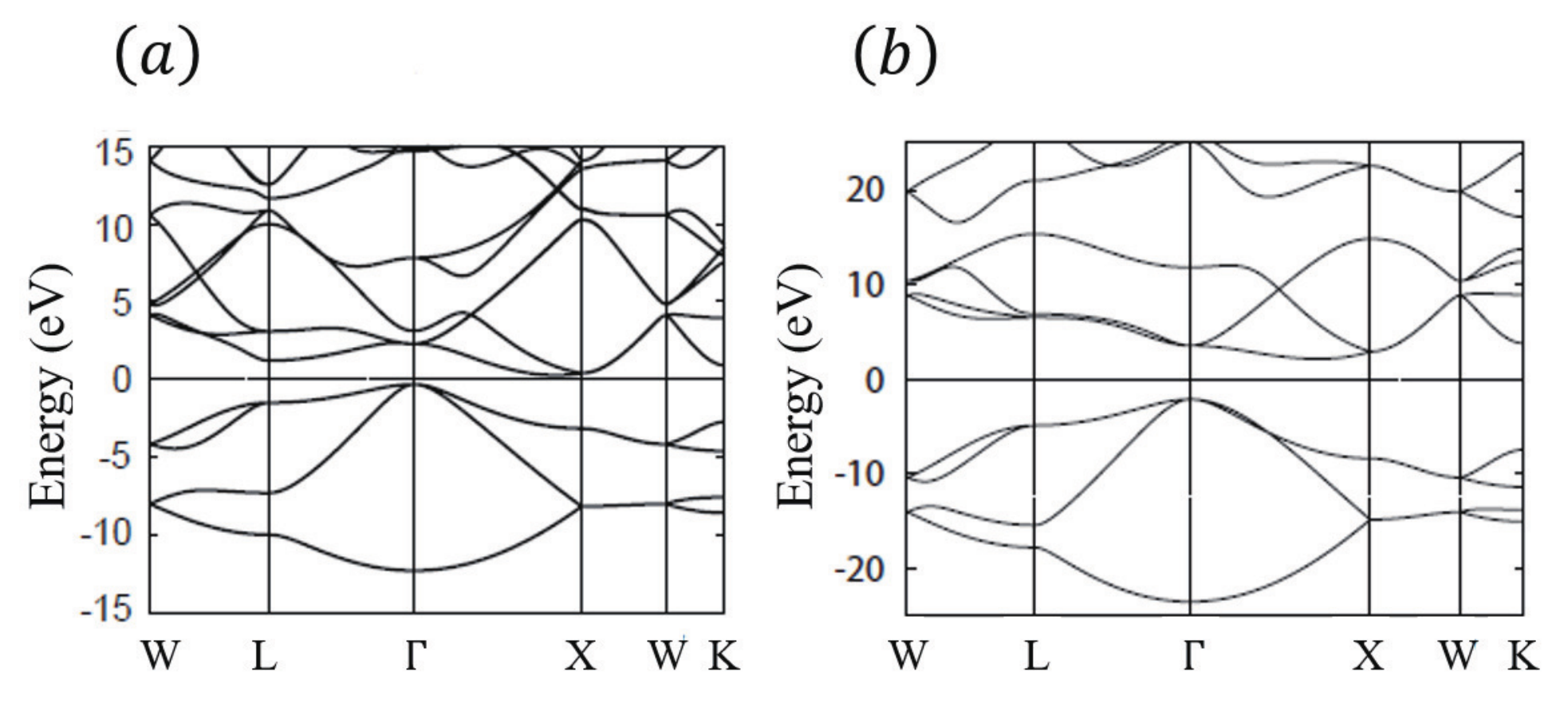}
 \caption{Band structures of (a) diamond and (b) silicon.}
 \label{fig:BulkEnergyBand_DiamondAndsilicon}
\end{figure}

In the following, we show how the topological surface states in diamond and silicon appear, 
based on the generalized gradient approximation (GGA) of the density functional theory (DFT)
and discuss physical origins for the topological nature of the bands. 
The computational condition is shown in Appendix \ref{appB}.
Their band structures are shown in Fig.~\ref{fig:BulkEnergyBand_DiamondAndsilicon}.
Although diamond has a larger gap than silicon,
both have qualitatively the same band structure.
The valence and conduction bands near the Fermi level originate from $sp_3$ hybrid orbitals.
The energy gap originates from covalent bonds of the $sp_3$ hybrid orbitals.

%%%TABLE2%%%%
  \begin{table}[tbp]
   \caption{Zak phase of diamond and silicon. The Zak phase is $\pi$ along [111] and along [11$\bar{1}$]}
   \begin{tabular}{|c|c|c|} \hline $\bm{n}$
    & Zak phase & \shortstack{number of dangling bond}  \\ \hline
    [111] & $\pi$ & $1$\\ \relax
    [11$\bar{1}$] & $\pi$ & $3 $ \\ \relax
    [001] &$ 0$ & $2$ \\ \relax
    [110] &$ 0$ & $2$ \\ \hline
   \end{tabular}\label{Ab:ZakPhase}
  \end{table}
%%%%%%%%%%%%%%%

%%%FIG13%%%%
%%%%%%%%%%

%%%FIG14%%%%
    \begin{figure*}[tbp]
    \centering
    \includegraphics[width=160mm]{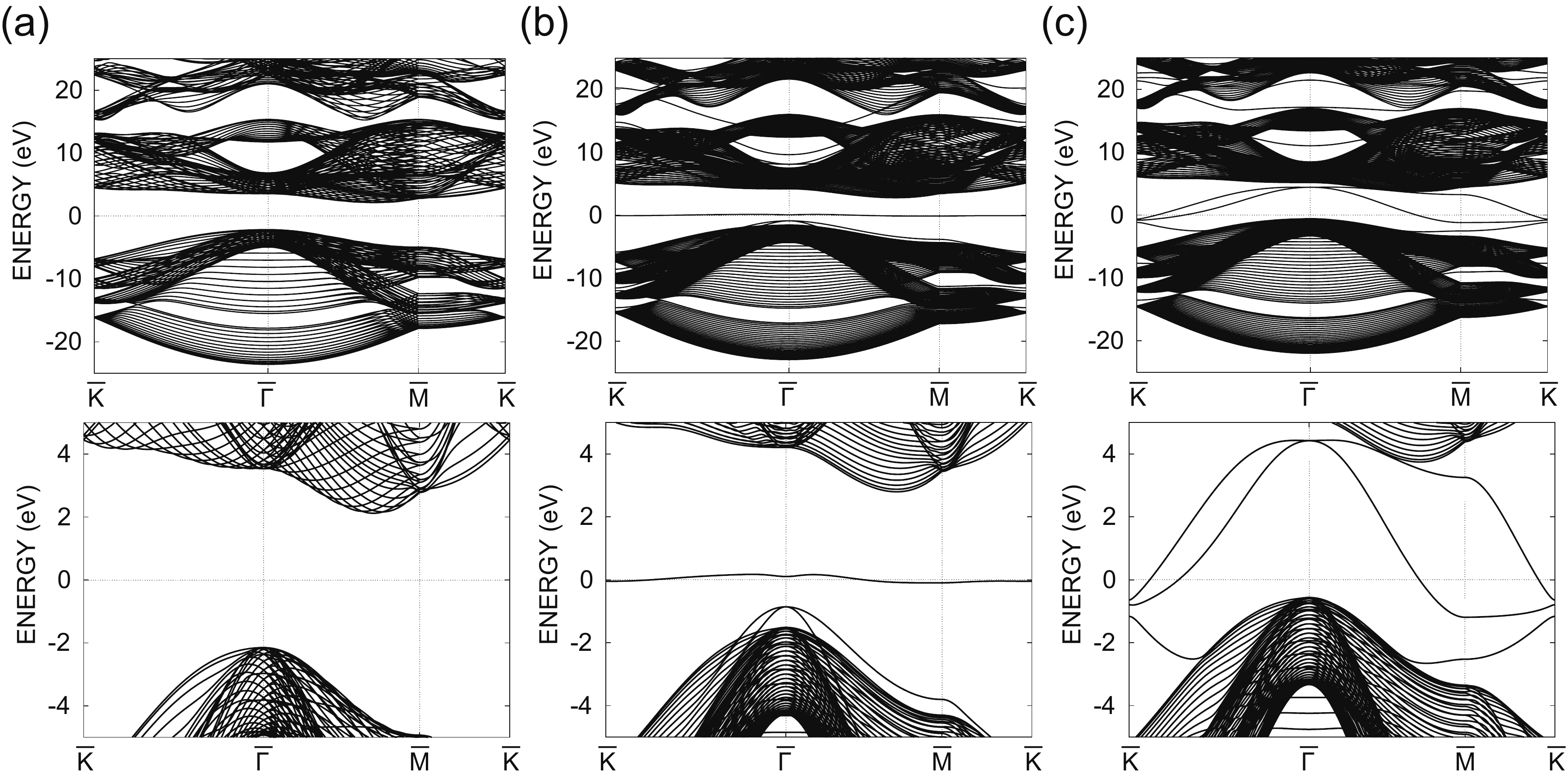}
    \caption{Band structures for a slab of diamond along (111) plane. The unit cell of the slab contains 60 atoms. (a) Slab with periodic boundary condition. The system has no surface, and therefore surface states do not appear. (b) Slab with (111) surfaces. There is one dangling bond per surface unit cell, and there are two degenerated surface states in the gap.  The dispersion of the surface states is almost flat. (c) Slab with (11$\overline{\rm 1}$) surfaces. There are three dangling bonds per surface unit cell, and there are six surface states in the gap. They are doubly degenerate. Dispersion of edge states in this direction is large compared with that in the  (111) direction in (b).}
   \label{fig:BandStructure_Diamond_Slab}
\end{figure*}
%%%%%%
%%%FIG15%%%
\begin{figure*}[tbp]
 \centering
  \includegraphics[width=130mm]{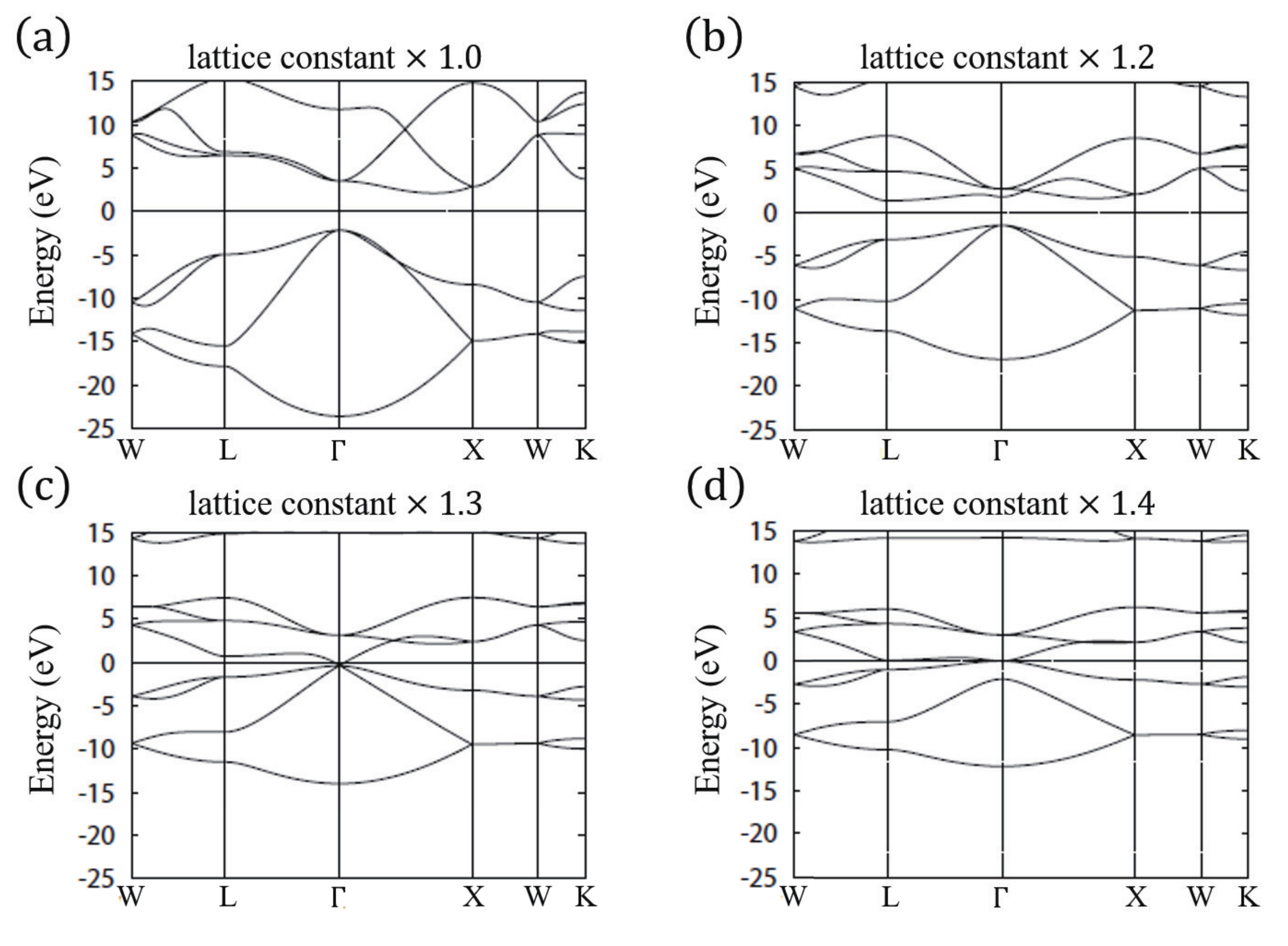}
 \caption{Changes in the band structure of diamond with increasing the lattice constant. We set the lattice constant to be the original one times (a) $1.0$, (b) $1.2$, (c) $1.3$ and (d) 1.4. The gap closing and band inversion occurs at $\Gamma$ in (c). }
 \label{fig:AtmicLimit}
\end{figure*}
%%%%%%%%%%

As we explain here, the Zak phase is related with dangling bonds at the surface.
The Zak phases of these materials along [111], [11$\overline{\rm 1}$], [001] and [110] directions are calculated by {\it ab intio} calculation. Table~\ref{Ab:ZakPhase} shows the calculation results of the Zak phase for diamond and silicon.  It shows that the Zak phases along [111] and along [11$\overline{\rm 1}$] directions are $\pi$ modulo $2\pi$, while those along [001] and along [110] directions are $0$ modulo $2\pi$. In these covalent crystals, 
the value of the Zak phase $\theta=\pi$ ($\theta=0$) coincides with an odd number (an even number) 
of dangling bonds per surface unit cell on each surface. This is because one dangling bond per surface unit cell supports one surface mode, which adds $\pi$ to the Zak phase. In the following, we confirm this for the diamond lattice. The energy bands of diamond slabs with (111) and (11$\overline{\rm 1}$) surfaces are shown in Fig.~\ref{fig:BandStructure_Diamond_Slab}. Figure~ \ref{fig:BandStructure_Diamond_Slab} (a) shows the band structure when a periodic boundary condition is imposed. Here, dangling bonds do not exist and there is no surface state. On the other hand, Fig.~\ref{fig:BandStructure_Diamond_Slab} (b) shows the band structure for a slab with (111) surfaces. There is a dangling bond per unit cell, supporting
one surface mode on each surface,
Thus the two surfaces of the slab support two surface modes, as  Fig.~\ref{fig:BandStructure_Diamond_Slab} (b) shows.  On the other hand, for the (11$\bar{1}$) surfaces (see Fig.~\ref{fig:BandStructure_Diamond_Slab} (c)), there are three dangling bonds per unit cell on each surface, and as a result, there exist six surface modes in the gap.

Here, we discuss  flatness of surface-state dispersions and its relation to anomalous dielectric response. As we discussed in Sec.~VA, the models discussed in Secs.~II-IV have dispersionless topological surface states due to chiral symmetry, and
an anomalous dielectric response is expected. 
In contrast, real materials do not have chiral symmetry. As a result, in silicon and diamond, 
the surface states protected by the $\pi$ Zak phase have dispersions. 
In such cases, the two surface modes on the two surfaces are half filled and are metallic. 
Under a small electric field, the surface remains metallic; in other words, there is still an overlap in energy between the surface states on one surface and
those on the other surface. Therefore, an abrupt charge transfer does not take place, and the polarization does not jump. 
The strength of the electric field has a threshold fro a charge transfer from one surface to the other, 
leading to the uprise of the polarization. In silicon, we see from 
Figs.~\ref{fig:BandStructure_Diamond_Slab} (b)(c) that 
the band structure of the edge states are almost flat in the slab with (111) surfaces, while far from flat with (11$\bar{1}$) surfaces. 
Therefore, the anomalous dielectric response is expected in the slab with (111) surfaces. 

Since any system in the atomic limit is always in a topologically trivial phase, topological phase transition should occur when the interatomic distance in diamond is extended from the topological insulator phase with a $\pi$ Zak phase. We show the change of band structure in diamond with various lattice constant in Fig.~\ref{fig:AtmicLimit}.  We set the lattice constant to be the original one times $1.0, 1.2, 1.3$ and $1.4$. These figures show that the band inversion occurs when the lattice constant is extended by $1.3$ times. If we further enlarge the lattice constant, it becomes a filling-enforced semimetal, because threefold degenerate states at the $\Gamma$ point are at the Fermi energy, and they bridge between the valence and conduction bands.

%%%
\subsection{Surface reconstruction}
Thus far, we have ignored the surface reconstruction. Since the electron energy in a dangling bond is higher than that in the valence band, the atoms on the surface are generally displaced so as to reduce the number of dangling bonds. It is called surface reconstruction. 
The  effect of surface reconstruction has been discussed in Sec.~IIID in Ref.~\cite{Vanderbilt2};
when the surface unit cell is multiplied by $M$ times by surface reconstruction, where $M$ is an integer, the polarization 
is defined modulo $e/(MS)$ instead of $e/S$.
Here we discuss the Zak phase and resulting topological surface states after surface reconstruction. It is well known that the silicon (111) surface forms $7\times7$ structure \cite{schlier1959structure}. In this structure, $49$ original surface unit cells form one new unit cell. The Zak phase is multiplied by 49 times, because the surface Brillouin zone is folded down by a factor of $1/49$. We note that the integrand in the formula of the Zak phase remains the same because the Zak phase is a bulk quantity independent of the details of the surface. 
Thus as a result the Zak phase
changes from $\theta=\pi$ to $\theta=49\pi\equiv\pi$ modulo $2\pi$, and the bulk polarization for a weak electric field 
is expected to be $ e/(2S\times49)=e/(98S)$ mod $e/(49S)$ per 
original surface unit cell. Here $49S$ is the area of the surface superstructure. In general, when $M$ unit cells form a superstructure, the Zak phase $\theta$ is given by
\begin{align}
 \theta &\equiv M\pi\equiv
\begin{cases}
\pi~(\text{mod}\ 2\pi)  & (\text{$M =$ odd}),\\
0~(\text{mod}\ 2\pi) & (\text{$M=$ even}).
\end{cases}
\end{align}
Here, $\theta$ does not depend on the wave vector $\bm k_{\parallel}$ because we consider only insulators. 
This result agrees with a counting argument of dangling bonds. When $M=$ even, all dangling bonds form pairs with covalent bonds, and there are no surface states in the gap, and the Zak phase becomes zero. Thus, the dielectric response is normal. Meanwhile when $M$ is odd, the electric polarization converges to $P=e/(2MS)~\text{mod}\ e/(MS)$, where $MS$ is a surface area of the superstructure. 

The previous studies show the correspondence between the survival of surface states and the parity of the number of unit cells constituting the superstructure in silicon \cite{takagi2008topologically,smeu2012electronic}.
In particular in Ref.~\onlinecite{smeu2012electronic}, the density of states of the silicon (111) surfaces with surface reconstuctions $2\times2$, $\sqrt{3}\times\sqrt{3}$, $5\times 5$ and $7\times 7$ 
is shown. Among these four cases, only for the $2\times2$ surface, where $M=2\times 2=4$ is even, the surface states have a gap, whereas in the other three cases with odd $M$,
the surface is metallic, suggesting existence of half-filled midgap surface states. These results are in perfect agreement with our results in this paper. 
  
%%%FIG16%%%%
\begin{figure*}[tbp]
	\centering
	\includegraphics[width=150mm]{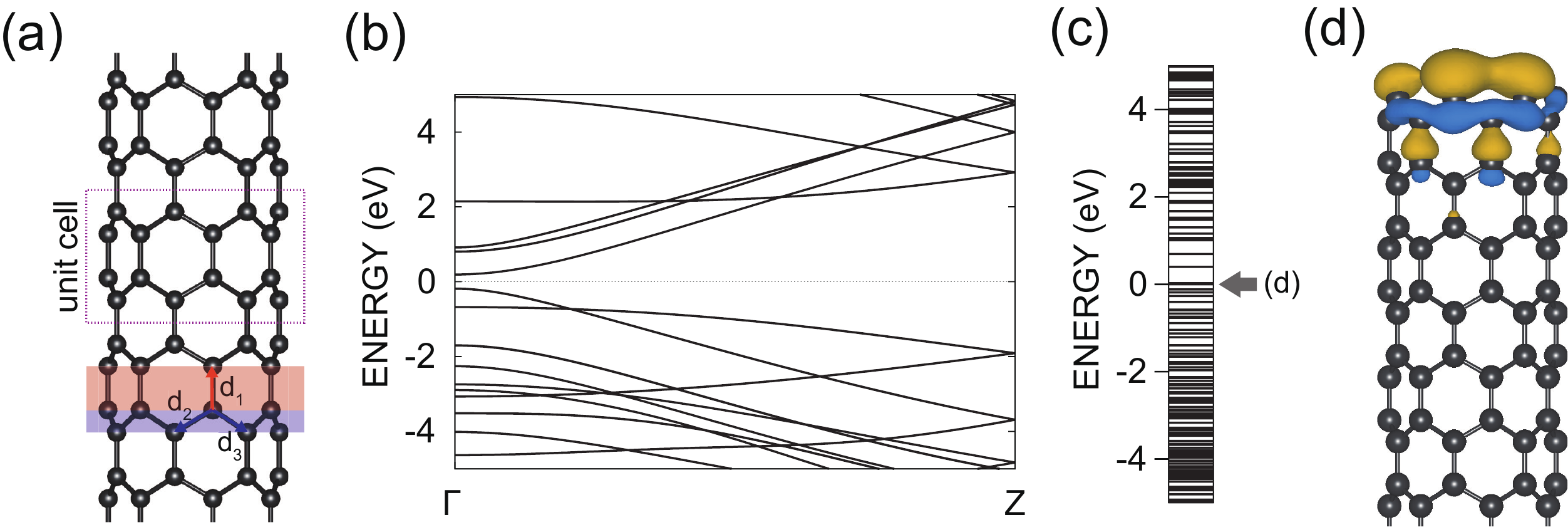}
	\caption{Topological edge states in a zigzag carbon nanotube. 
		(a) Crystal structure of the zigzag carbon nanotube with the (7,0) chirality.
		(b) Electronic band structure of the (7,0) nanotube. The Fermi energy is at $E=0$. 
		(c) Eigen energies of the (7,0) nanotube with a finite length, containing 10 unit cells, i.e. 280 carbon atoms.		
		(d) Distribution of the electron density for the eigenstate at the Fermi level of the (7,0) nanotube with finite length.
It is localized at the edge of the nanotube.	
	}
	\label{fig:BandStructure_Nanotube}
\end{figure*}
%%%%%%  

\subsection{Carbon nanotube}

Next, we show the result of a carbon nanotube.
%Carbon nanotubes have inversion symmetry in the case of zigzag and armchairs.
While a carbon nanotube of the armchair type is always semimetallic, that of the zigzag type has a gap when the chirality is $(3i+1, 0)$ or $(3i+2, 0)$ $(i=1,2,...)$. As we limit ourselves on insulating systems, we focus on the $(m,0)$ zigzag nanotube, with $m$ not an integer multiple of three. 
As the number of dangling bonds in this nanotube is $m$, we restrict ourselves to the case with odd $m$, so that the Zak phase is
$\pi$.
Figure~\ref{fig:BandStructure_Nanotube}(a) shows a structure of zigzag nanotube with a chirality of (7,0).
The $p_z$ orbitals forming the $\pi$ bonds have a small gap around the Fermi level (Fig.~\ref{fig:BandStructure_Nanotube}(b)).
Because the blue region in Fig.~\ref{fig:BandStructure_Nanotube}(a) has about twice as much transfer as the red region, the Wannier center of the $\pi$ covalent bond are located at the blue region.
Therefore, the $p_z$ orbital forming the $\pi$ bonds has no polarization in the unit cell corresponding to the zigzag edge.
On the other hand, since $\sigma$ covalent bonds are cut at the $(m, 0)$ zigzag edge, the Zak phase is $\pi$ when $m$ is an odd integer.
Figure~\ref{fig:BandStructure_Nanotube}(c) shows the eigen energies of a finite nanotube with 10 unit cells in length.
Half-filled edge states exist at the Fermi level.
These are topological states originated from the $\pi$ Zak phase and are localized at the edges (Fig.~\ref{fig:BandStructure_Nanotube}(d)).

\subsection{Topological electrides}
In our previous work \cite{hirayama2018electrides}, we proposed that the electrides can be good candidates for topological materials. Among such topological electrides, we  
proposed that Sc$_2$C is an insulator with $\pi$ Zak phase along the [111] direction. 
It is a layered material,
and is a two-dimensional eletride, where the Wannier orbitals reside between the atomic layers along the (111) plane. It can be cleaved along the (111) plane, which cuts the Wannier orbitals 
into half. This is similar to the SSH model on the topological regime, and it leads to in-gap surface states, which is a remnant of the electronic states between the atomic layers. 
Since existence of electronic states at interstitial regions is a hallmark of the electrides, this characteristic of the electrides is the key to realize the surface with the $\pi$ Zak phase. 

The $\pi$ Zak phase in this topological electride shows that the notion of the Zak phase 
has wider applicability than counting of dangling bonds at the surface. 
In Sec VIA, we have discussed that in covalent crystals, the number of dangling bonds being even and odd corresponds 
to the Zak phase 0 or $\pi$, respectively. 
On the other hand, in Sc$_2$C, we can make
the (111) surface without appearance of dangling bonds. Even in such crystals we can
adopt the notion of the Zak phase, and indeed we can see appearance of midgap topological 
surface states on the (111) surface of Sc$_2$C.

\subsection{\textit{Ab initio} calculation for dielectric response}

%%%FIG XXX %%%%
\begin{figure}[tbp]
	\centering
	\includegraphics[width=75mm]{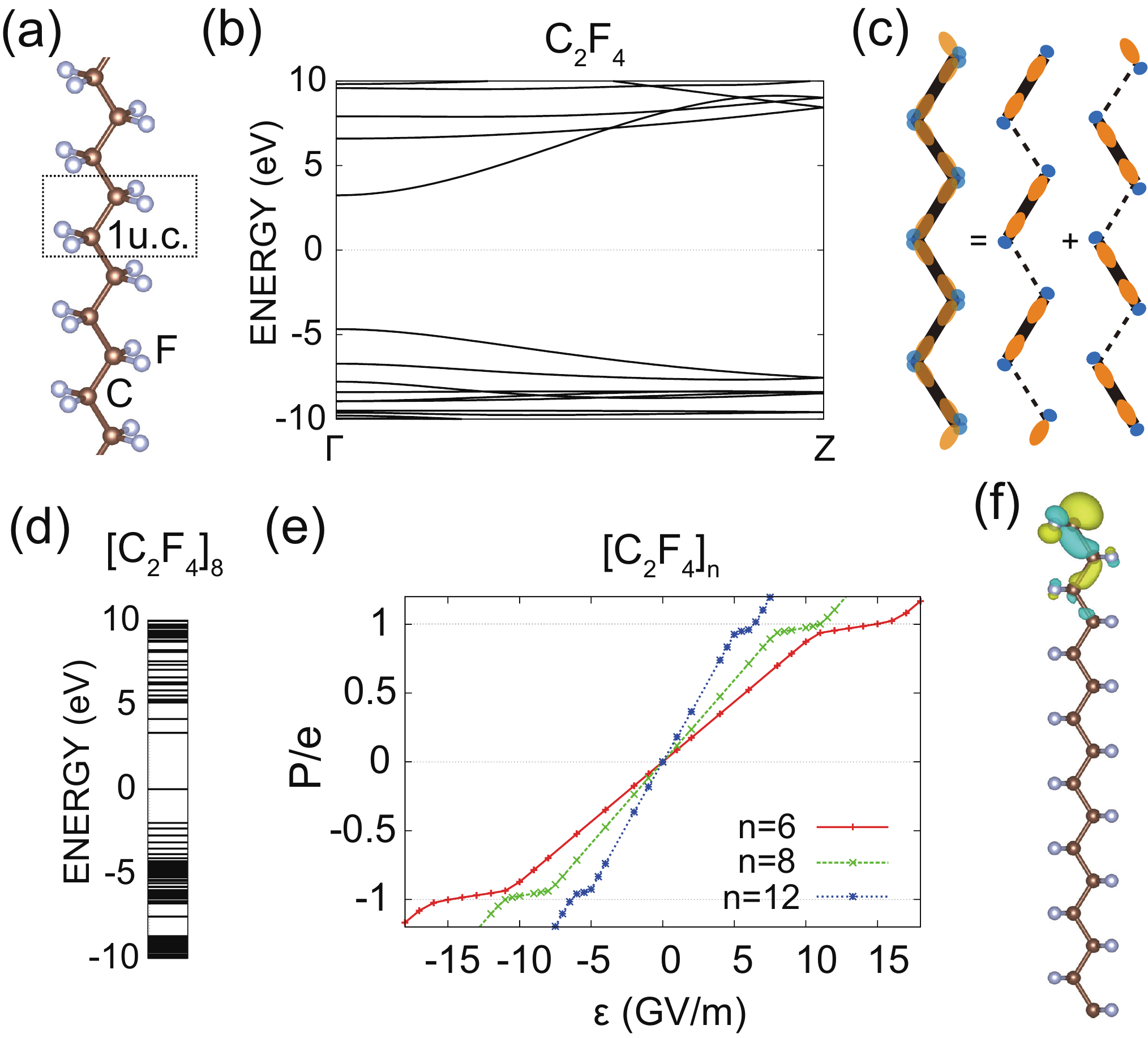}
	\caption{Dielectric response of the PTFE.
		(a) Structure of the PTFE.
		(b) Electronic band structure of the PTFE.
		The Fermi energy is at $0$.
		(c) Covalent bonds of the PTFE.
		(d) Enegy eigenvalues of [C$_2$F$_4$]$_{8}$.
		(e) Dielectric response of the finite PTFE.
		$P$ is the polarization.
		(f) LUMO of [C$_2$F$_4$]$_{8}$ under the electric field $\varepsilon= 9$ (GV/m). 
	}
	\label{fig:C2F4}
\end{figure}
%%%%%%  

Here we show the \textit{ab initio} calculation for the dielectric response.
First, we calculate the dielectric response of PTFE [C$_2$F$_4$]$_n$.
Figure~\ref{fig:C2F4}(a) shows the crystal structure of the PTFE.
The one-dimensional band structure of the PTFE has a large band gap (Fig.~\ref{fig:C2F4}(b)).
Such a large gap is suitable for the anomalous dielectric response (see Figs.~\ref{fig:fig3-4}(a)(b)). 
The C atoms are connected by the $\sigma$-bonds originating from the $sp^3$ orbitals  (Fig.~\ref{fig:C2F4}(c)).
%The covalent bond along the C atoms can be written by an extended SSH model with two orbitals for each site.
%Since the two orbitals form dimers at different positions 
Because the two orbitals per each site form covalent bondings, this system
is regarded as a superposition of two SSH models with dimers at 
different positions, 
as shown in Fig~\ref{fig:C2F4}(c). Therefore, the Zak phase is always $\pi$ regardless of the choice of the unit cell.

Next, we calculate the energy eigenvalues of a finite system of the PTFE (Fig.~\ref{fig:C2F4}(d)).
The topological edge states originating from the $\pi$ Zak phase exist at the Fermi level ($E=0$ in Fig.~\ref{fig:C2F4}(d)) .
Next we apply the electric field along the C chain to finite systems and calculate the polarization (Fig~\ref{fig:C2F4}(e)). 
We here discuss a condition for the system length to exhibit the anomalous dielectric response. 
If the length of the PTFE is too short, the hybridization between edges become large, which makes the dielectric response less sharp (Fig.~\ref{fig:fig3-4}(c)).
On the other hand, if the length of the PTFE is too long, the energy loss of the Coulomb interaction causes a gradual charge transfer in the system to locally satisfy charge neutrality.
Therefore, in a realistic system with a finite bulk gap,
the length of the system should be moderate in order to exhibit the anomalous response.
From the above reasons, we calculate the dielectric response of [C$_2$F$_4$]$_6$, [C$_2$F$_4$]$_8$, and [C$_2$F$_4$]$_{12}$ in Fig~\ref{fig:C2F4}(e), 
where the magnitude of ${\bm P}$ is shown. 
%The polarization ${\bm P}$ is estimated from the center of the finite system
%and we plot $|{\bm P}|$ and $-|{\bm P}|$ for positive and negative electric fields, respectively.
The electric response becomes flat around $P\sim \pm e$ in all the three cases. It perfectly agrees with Fig.~\ref{fig:fig3-4}(d), considering the spin degeneracy ($=2$). Moreover, the slope of the dielectric response 
becomes sharp for a longer chain, which also agrees with our theory.
We show the lowest unoccupied molecular orbital (LUMO) of [C$_2$F$_4$]$_{8}$ under the electric field $\varepsilon= 9$ (GV/m) in Fig~\ref{fig:C2F4}(f), which corresponds to one of the topological edge states originating from the $\pi$ Zak phase.
When the electric field becomes even larger, the electrons in the bulk region 
participate in screening the polarization, and the dielectric response becomes linear again (see Fig.~\ref{fig:fig3-4}(d)).

We also calculate the dielectric response of a (111) diamond slab with $1\times1$ surface structure discussed in Sec.~\ref{Ab initio calculations for diamonds}.
We find that the polarization does not show an anomalous response but has a linear dependence on the electric field.
We attribute the absence of the anomalous behavior of the dielectric response to the weak ${\bm k}$ dependence of 
the topological surface states in diamond (111) surface with $1\times1$ structure (Fig.~\ref{fig:BandStructure_Diamond_Slab}(b)), 
as discussed in Sec.~\ref{sec:dispersion}.
%Such a ${\bm k}$ dependence gives rise to a continuous change of polarization against the electric field.
On the other hand, in the silicon with realistic $7 \times 7$ surface reconstruction, the anomalous dielectri response is more likely because of the two reasons.
One is because the topological surface bands are almost flat, and 
the other is because the larger unit cell leads to the smaller energy increase of Coulomb repulsion per two-dimensional unit cell associated with the charge transfer, making the bulk region less likely to participate in the anomalous response.
Therefore, nano-sized silicon with the surface reconstruction would be promising to realize the anomalous response.
%In addition, since the energy increase per two-dimensional unit cell due to the movement of electrons between the surfaces is reduced to $1/49$ of that without the reconstruction, 
%electrons in the bulk region are less likely to participate in the anomalous response.
%Nano size silicon with the surface reconstruction would be useful system to realize the anomalous response.

\section{Bulk physics versus surface physics} \label{sec:bulk-surface}
In this section, we discuss how the anomalous dielectric response is related with the bulk polarization,
described in the modern theory of polarization. 
In the modern theory of polarization \cite{resta1992r,Vanderbilt1,Vanderbilt2,Resta2}, the notion of the 
bulk polarization is introduced, and it is formulated in term of the Zak phase, in the
absence of an electric field. A dielectric response to 
an electric field is also discussed along the same line \cite{NunesVanderbilt,Nunes,Combes}.
In these works, the dielectric response is studied as a bulk effect, which is considered 
to be valid in the thermodynamic limit.

We need some care in comparing our results with those 
in the previous works on bulk dielectric responses \cite{NunesVanderbilt,Nunes,SouzaIniguez,Umari,Combes}. 
In particular, the dielectric response is 
formulated by constructing 
the Wannier-Stark ladder perturbatively for a weak electric field in Ref.~\onlinecite{Combes}, and
a similar result on dielectric response is obtained in Ref.~\onlinecite{Swiecicki} as well.
In Ref.~\onlinecite{Combes}, this formalism is applied to the Rice-Mele model;
because the SSH model is a special case of the Rice-Mele model, 
we can directly compare those results with ours in this paper. 
As we explain later in detail, in the results in Ref.~\onlinecite{Combes} for the SSH model with the $\pi$ Zak phase, there 
is no divergence in the dielectric response $\chi\equiv \frac{dP}{d\varepsilon}$, which seems to contradict our result.

\begin{figure}[tbp]
	\centering
	\includegraphics[width=75mm]{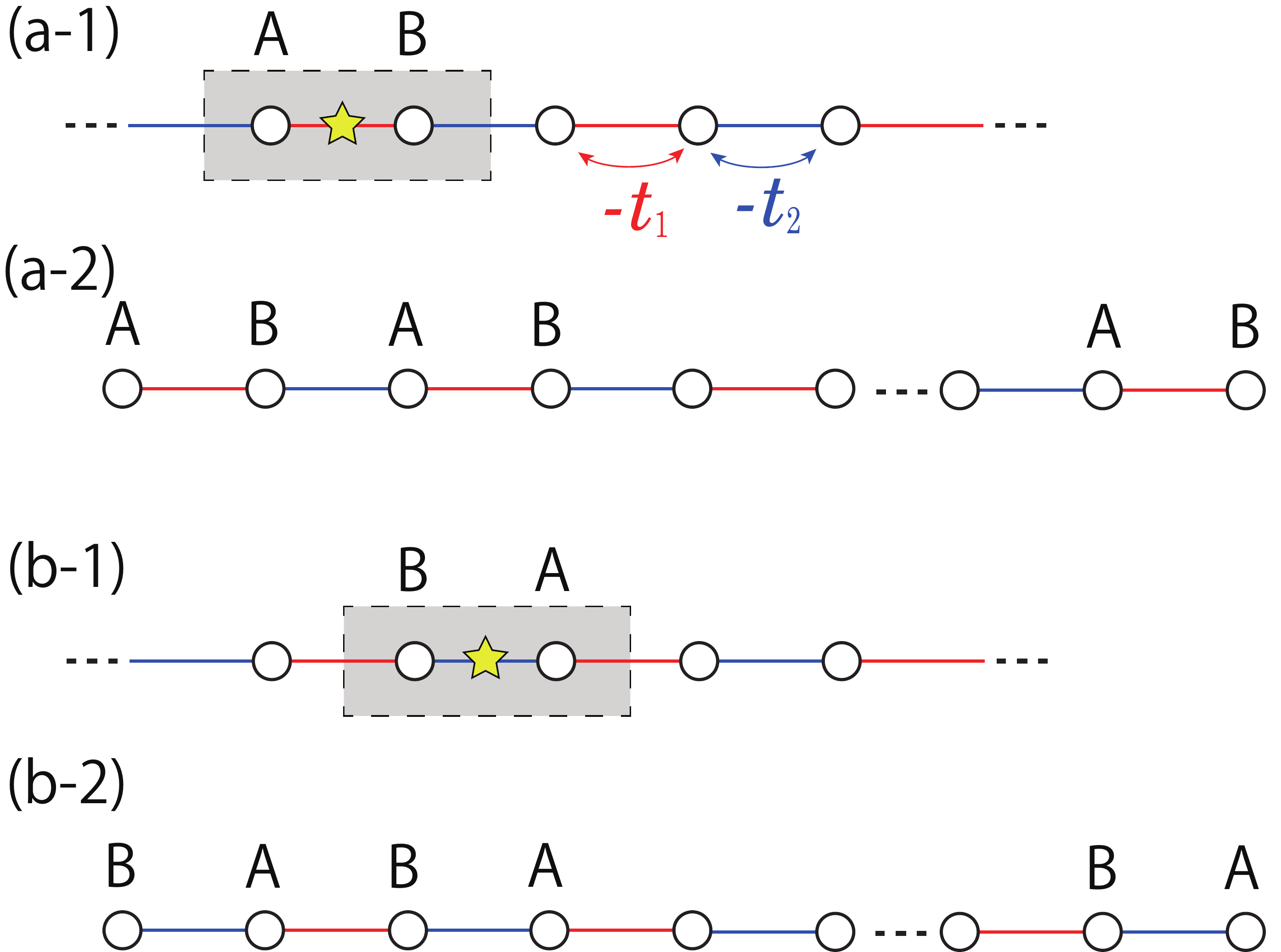}
	\caption{Choice of the unit cell and corresponding crystal termination in the SSH model.
		(a-1) and (b-1) show two choices of unit cell shown as a grey box.
		(a-2) and (b-2) are the SSH chain with terminations corresponding to (a-1) and (b-1), 
respectively. In each case, the star represents the corresponding inversion center.
	}
	\label{fig:unitcell}
\end{figure}

This discrepancy is related with an interplay between the bulk and surface physics. To explain this,
we focus on degrees of freedom in the choices of the crystal terminations and those of the unit cell 
\cite{Vanderbilt2,Chen-Lee}. Let us consider 1D systems with inversion symmetry, such as the SSH model in Sec.~II. In this case, as mentioned earlier, the unit cell should be invariant
under inversion symmetry. There are two choices of the inversion center, displaced by 
a half of the primitive translation vector, and correspondingly, we have two choices of unit cells. It is shown in Figs.~\ref{fig:unitcell} (a-1) and (b-1). This correspondence between the inversion center 
and the unit cell is required for the relationship between the Zak phase and inversion parities in Eq.~(\ref{eq:parity}).
By
changing the choice of the unit cell, the parity at $k=\pi$ changes sign, and the value of the Zak phase is switched between $0$ and $\pi$. In Ref.~\onlinecite{Vanderbilt2}, it is discussed that the 
crystal termination is related to the choice of the unit cell so that the unit cell, with displacement by
crystal translation vectors, should fully cover the whole system with surface termination. 
Thus, there are two ways of terminations for the SSH model, shown in Figs.~\ref{fig:unitcell} (a-2) and (b-2),
and they correspond to the choices of the unit cells shown in
Figs.~\ref{fig:unitcell} (a-1) and (b-1), respectively.  By employing this correspondence, 
the relationship between the $0$ ($\pi$) Zak phase and absence (presence) of the in-gap boundary 
states is guaranteed.

We now turn to 
the Rice-Mele model studied in Refs.~\onlinecite{Combes,Vanderbilt2}. In 
Ref.~\onlinecite{Combes}, this model is defined on a
lattice in Fig.~\ref{fig:fig1-2}(a), and is described by a $2\times 2$ Bloch Hamiltonian
\begin{equation}
H(k)=\left(\begin{array}{cc}
\Delta& 2t(\cos\frac{k}{2}-i\delta\sin\frac{k}{2})\\
2t(\cos\frac{k}{2}+i\delta\sin\frac{k}{2})&
-\Delta
\end{array}\right)
\end{equation}  
in the basis of A and B sublattices, 
where $t(1\pm \delta)$ represents the alternating hopping, and $\Delta$ is a staggered on-site
potential. If we start from our SSH model in Eq.~(\ref{SSH:BulkHamil}), this Rice-Mele model 
is realized by putting 
$t_1=-t(1+\delta)$, $t_2=-t(1-\delta)$, and by adding a staggered on-site potential $\pm\Delta$ 
at the A and B sublattices. 
In Ref.~\onlinecite{Combes}, the spontaneous magnetization and susceptibility is studied for the
values of parameters $t=1$, $\Delta=0.6\cos\theta$ and $\delta=0.6\sin\theta$ with a parameter $\theta$. 
We note that $\theta=3\pi/2$ corresponds to the SSH model with the $\pi$ Zak phase. 
It is found that the  dielectric response $\chi\equiv \frac{dP}{d\varepsilon}$ has no singularity 
even at $\theta=3\pi/2$, and that the
bulk spontaneous polarization, defined modulo $e$, has no anomaly at $\theta=3\pi/2$. 
It means that there is no anomalous dielectric response in the bulk polarization even at $\theta=3\pi/2$.
On the other hand, 
the polarization for a finite chain is also calculated in Refs.~\onlinecite{Combes,Vanderbilt2}, and it
has a jump 
between $e/2$ and $-e/2$ at 
$\theta=3\pi/2$. This implies that the system is at the verge between the states with polarizations
$+e/2$ and $-e/2$, and that a tiny electric field can easily drive the system to obtain polarization $\pm e/2$.
Thus, the bulk polarization is defined only modulo $e$, while the polarization for a finite chain
is not modulo $e$. The latter can be determined without ambiguity once the finite-length chain, including its
terminations, is fixed. The bulk polarization and the polarization for a finite chain 
are identical only modulo $e$ (if we ignore small finite-size effect, which converges to
zero for a long-chain limit). Thus the jump of the polarization between $-e/2$ and $e/2$ 
never appears in the bulk polarization, because of its ``mod $e$'' nature. 
In other words, in the SSH model, the anomalous dielectric response discussed in the 
present paper depends on the terminations, and in the case $|t_2|>|t_1|$, it appears only in the termination in 
Fig.~\ref{fig:unitcell}(a-2) but not in the termination in 
Fig.~\ref{fig:unitcell}(b-2). Thus it is natural that the bulk polarization does not 
have a singularity for the SSH model with the $\pi$ Zak phase (i.e. $\theta=3\pi/2$), 
where we have shown the anomalous dielectric response.

%%%FIG16%%%%
%%%%%%  

\section{Conclusion}\label{Conclusion}

To summarize, we have shown anomalous dielectric response in insulators with the $\pi$ Zak phase. We demonstrated it in various tight-binding models in one, two and three dimensions.  In each model,  the polarization suddenly rises to a large value close to $e/2$ per edge or surface unit cell by application of an external electric field when the Zak phase is $\pi$ over the entire Brillouin zone. By using effective models, we have confirmed that the topological surface states protected by the $\pi$ Zak phase give rise to this phenomenon. We also show, by using {\it ab-intio} calculation, that diamond and silicon slabs with (111) surfaces have the $\pi$ Zak phase over the entire Brillouin zone and relatively flat midgap states. Our results are expected to be applicable to these materials even when surface reconstruction occurs. We also 
show such a dielectric response of the PTFE  with plateaus at $P\sim \pm e$ by {\it ab initio} calculation, in agreement with our theory. 

We have also discussed that this anomalous dielectric response do not
appear for the bulk polarization described by the modern theory of polarization, because 
this effect depends on the termination of the system, and therefore is attributed to the surface effect.

\begin{acknowledgments}
The authors gratefully acknowledge useful discussions
with D.~Vanderbilt.
This work was supported by JSPS KAKENHI
Grant No.~JP18H03678, by MEXT KAKENHI
Grant No.~JP20H04633, by CREST, JST (No.~JP-MJCR14F1),
and by the MEXT Elements Strategy Initiative to Form
Core Research Center (TIES), Grant Number JPMXP0112101001.
\end{acknowledgments}
\appendix

%%%%%%%%%%%%%%%%%%%%%%%%%%%%%%%%%%%%%%%%
\section{ Two-dimensional models (Models III and IV)}\label{2D}
%%%%%%%%%%%%%%%%%%%%%%%%%%%%%%%%%%%%%%%%

In Sec.~II,  we have seen that the dielectric polarization in the SSH model abruptly increases up to $\pm e/2$ in response to a weak electric field when the Zak phase is $\pi$. We expect similar phenomena in two-dimensional insulators. To confirm that, we consider two tight-binding models on the honeycomb lattice: the Model III with anisotropic hopping amplitude and the Model IV with three orbitals per lattice site. 
In both models, the lattice consists of two sublattices A and B. Let $a$ denote the distance between the two nearest A sites, and we set the nearest bond vectors from the A site to the adjacent B sites $\bm d_{i=1,2,3}$ to be $\bm d_1=\frac{a}{\sqrt{3}}(0,1), \bm d_2=\frac{a}{\sqrt{3}}(-\frac{\sqrt{3}}{2},-\frac{1}{2}), \bm d_3=\frac{a}{\sqrt{3}}(\frac{\sqrt{3}}{2},-\frac{1}{2})$ as shown in Fig. \ref{fig:fig5-6}(a). 
We note that in two-dimensional systems, the polarization, which is an electric dipole moment per unit area, has the unit of charge divided by a length.

%%%
\subsection{Zak phase and dielectric polarization for two-dimensional systems}
%%%
First, we review the Zak phase and its connection with electric polarization in two-dimensional insulators. 
The Zak phase with an integration path along a reciprocal lattice vector $\bm G_\perp$ is defined as
\begin{align}
\theta(k_\parallel)=-i\sum_n^{\text{occ.}}\int_0^{|\bm G_\perp|}dk_\perp\bra{u_n(\bm k)}\dfrac{\partial}{\partial k_\perp}\ket{u_n(\bm k)},\label{2D:Zak phase}
\end{align}
where $\bm k =k_\perp \bm n +k_\parallel\bm l$ and $\bm n={\bm G}_{\perp}/|\bm G_{\perp}|, \bm n \perp \bm l, |{\bm l}|=1$. This Zak phase is related with a polarization in a ribbon geometry with its edges along $\bm l$, defined at $k_\parallel$ in the surface Brillouin zone. The Zak phase $\theta(k_\parallel)$ is quantized as 0 or $\pi$ modulo $2\pi$ under both the TR and SI symmetries \cite{kariyado2013symmetry}. Therefore, in insulators, this quantized value of $\theta(k_\parallel)$ is constant for any $k_\parallel$. In the region of the surface Brillouin zone where $\theta(k_\parallel)=\pi$, there exist degenerated edge states at zero energy when the system has chiral symmetry \cite{ryu2002topological}. The electric polarization $P_0$ across the ribbon can be obtained from 
\begin{align}
P_0=-e\int_0^{2\pi/a} \dfrac{d k_\parallel}{(2\pi)^2}\theta(k_\parallel)\ ({\rm mod}\ e/a),
\end{align}
%where $a$ is the length of the edge unit cell.
where $a$ is the lattice constant along the edge parallel to ${\bm l}$.
This equation indicates that the polarization takes a large value of $e/(2a)$ in the system with the $\pi$ Zak phase for any $k_\parallel$. This nonzero value of
the polarization $P_0$ does not contradict the inversion symmetry of the system, because $P\equiv e/(2a)\equiv -e/(2a)\ $(mod $e/a$). 

%%%%%%%%%%%

%%%%%%%%%%%%%%%%%%%%%%%%%%%%%%%%%%%%%%%%%%%%%%%%%%%%%%%%%%%%%%%%%%%
\subsection{Model III: Anisotropic tight-binding model on the honeycomb lattice}\label{MAHA}
%%%%%%%%%%%%%%%%%%%%%%%%%%%%%%%%%%%%%%%%%%%%%%%%%%%%%%%%%%%%%%%%%%%

\begin{figure}[tbp]
    \centering
    \includegraphics[width=86mm]{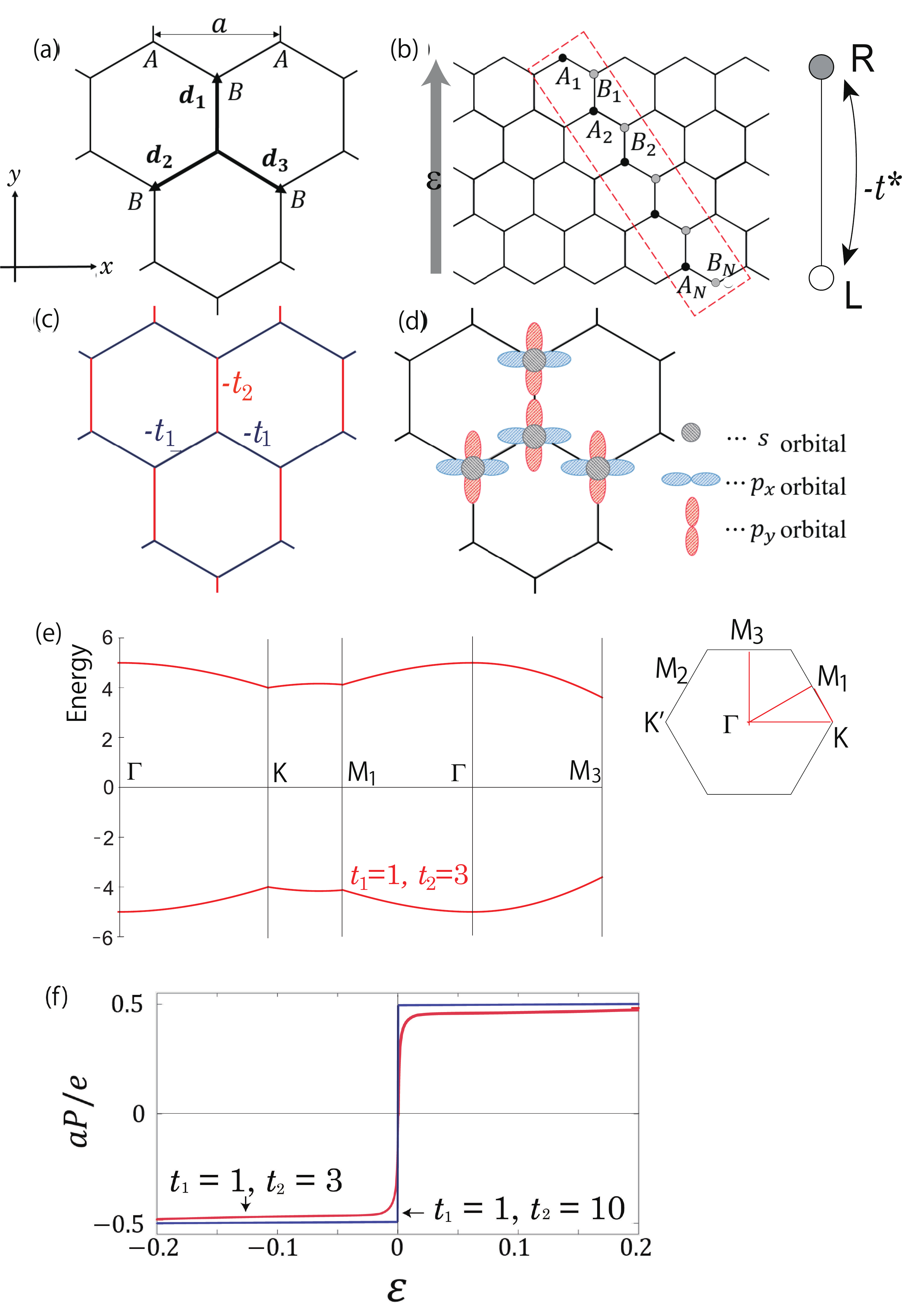}
    \caption{Two-dimensional models III and IV. (a) Honeycomb lattice. The lattice constant is $a$ and the nearest-neighbor bond vectors from the A site to the B site are $\bm d_{i=1,2,3}$. (b) Zigzag-edge nano ribbon. The red dotted line shows the unit cell in the ribbon. (c) Schematic figure of the model with anisotropic hopping amplitudes (Model III). (d) Schematic figure of the model with three orbitals per lattice site (Model IV). (e) Band structure for the Model II with $t_1=1$, $t_2=3$. The bands are shown along high-symmetry lines shown in red in the two-dimensional Brillouin zone in the right panel. (f) Dielectric polarization $P(\varepsilon)$ of the model III in a nano-ribbon with zigzag edges. $P(\varepsilon)$ rapidly changes toward $\pm e/2a$ in the vicinity of $\varepsilon=0$. Here we set $N=10$ and the number of sites in the unit cell is $20$. The hopping amplitudes $(t_1,t_2)$ are  $(1,3)$ for the red line and (1,10) for the  blue line. As the anisotropy increases, the change in the polarization becomes sharper.}
   \label{fig:fig5-6}
\end{figure}

In this section, we consider an anisotropic tight-binding model on the honeycomb lattice. The  hopping amplitudes between the nearest-neighbor sites are anisotropic. The hopping amplitude along $\bm d_1$ is $-t_2$ and those along other directions are $-t_1$, as shown in Fig.~\ref{fig:fig5-6} (c). $t_1$ and  $t_2$ are set to be positive. The Hamiltonian is given by
\begin{align}
H_0=-\sum_{\langle ij\rangle}t_{ij}c^{\dagger}_ic_j\ ,\label{HamilOfNonEqHon}
\end{align}
where $t_{ij}$ is the hopping amplitude from site $i$ to $j$. First, the bulk Hamiltonian is given by
\begin{align}
H^{\text{bulk}}(\bm k)&=\left(\begin{array}{cc}
0&R(\bm k)\\
R^\ast(\bm k)&0
\end{array}\right)\\\
%R(\bm k)&= t_1e^{-i{\bm k}\cdot{\bm d_1}}+t_2(e^{-{\bm k}\cdot {\bm d_2}}+e^{-i{\bm k}\cdot {\bm d_3}}).
R(\bm k)&=-t_2e^{-i{\bm k}\cdot{\bm d_1}}-t_1(e^{-{\bm k}\cdot {\bm d_2}}+e^{-i{\bm k}\cdot {\bm d_3}}).
\end{align}
The energy eigenvalues are given by 
\begin{align}
E_{\bm k}^2&=\left(t_2+2t_1\cos\frac{k_xa}{2}\cos\frac{\sqrt{3}k_ya}{2}\right)^2\notag\\
&+4t_1^2\cos^2\frac{k_xa}{2}\sin^2\frac{\sqrt{3}k_ya}{2}.
\end{align}
A band gap exists when $t_2>2t_1$ (see Fig.~\ref{fig:fig5-6}(e)) \cite{dietl2008new,takahashi2013completely}. Thus, we can define the dielectric polarization only when $t_2>2t_1$. 
By taking $\bm n=(0,1)$ and $\bm l=(1,0)$, the Zak phase along the $y$-axis is given by
\begin{align}
\theta_y{(k_x)}=\begin{cases}
\pi \ \left(\cos k_xa<\dfrac{t_2}{2t_1}\right)\\
0 \ (\text{otherwise}) 
\end{cases}.
\label{eq:ZakmodelB}
\end{align}
When the band gap is open, i.e. $t_2>2t_1$, the Zak phase equals to $\pi$ in the whole Brillouin zone. The $\pi$ Zak phase represents the polarization $e/(2a)({\rm mod}\ e/a)$ along the $y$-axis  \cite{delplace2011zak}. 
Here, thanks to the SI symmetry, the Zak phase $\theta_y(k_x)$ with $k_x$ being one of the one-dimensional TRIM, i.e. $k_x=0$ and $k_x=\pi/a$, can be obtained from parity eigenvalues of the occupied states at TRIM.
Namely, 
\begin{align}
e^{i\theta_y(k_x=0)}&=\prod_n^{\text{occ.}}\xi_n(0,0)\xi_n\left(0,\frac{2\pi}{\sqrt{3}a}\right), \\
e^{i\theta_y\left(k_x=\frac{\pi}{a}\right)}&=\prod_n^{\text{occ.}}\xi_n\left(\frac{\pi}{a},\frac{\pi}{\sqrt{3}a}\right)\xi_n\left(\frac{\pi}{a},-\frac{\pi}{\sqrt{3}a}\right), 
\end{align}
where $\xi_n({\bm k})$ is the parity eigenvalue of the $n$th eigenstate. 
It reproduces the result in  Eq.~(\ref{eq:ZakmodelB}). 
Here,
the unit cell for the bulk used in the calculation of $\xi_n$ is taken as a pair of a A-site and a B-site, displaced by ${\bm d_2}$, so that the the unit cell covers the whole ribbon by translating the unit cell by the translation vectors, 
and the inversion center is located at the midpoint of the unit cell.

%%%FIG5%%%%
%%%%%%%%%%%

We next study the dielectric polarization for a ribbon with zigzag edges as shown in Figs. \ref{fig:fig5-6} (b) and (c).  There are $2N$ sites in the unit cell of the zigzag-edge nano ribbon. Let $\sl A_i$ and $\sl B_i$ $(i=1,2,\cdots,N)$ denote the $2N$ lattice sites in the unit cell as shown in Fig. \ref{fig:fig5-6} (b). The Hamiltonian of this system is given by
\begin{align}
H_0(k_x)&=-t_1\sum_{n=2}^{N}e^{-i\tilde d_1k_x}b^\dagger_{k_x,n-1}a_{k_x,n} \notag \\
&\ \  - t_2\sum_{n=1}^N\left(e^{-i\tilde d_2k_x}+ e^{-i\tilde d_3k_x}\right)b^{\dagger}_{k_x,n}a_{k_x,n}+{\sl h.c.} \label{HamRibAn}
\end{align}
where $a_{k_x,n}^\dagger(b_{k_x,n}^\dagger)$ and $a_{k_x,n}(b_{k_x,n})$ are the creation  and annihilation operators of an electron with Bloch wave vector $k_x$ at the $A_n$ ($B_n$) site in the unit cell, and $\tilde{d}_{i}~(i=1,2,3)$ is the $x$ component of $\bm d_{i}$. When an electric field of strength $\varepsilon$ along the $y$ axis is applied, the Hamiltonian is given by
\begin{align}
H(k_x)&=H_0(k_x)+H_\varepsilon(k_x),\\
H_\varepsilon(k_x)&=\dfrac{ae\varepsilon}{\sqrt{3}}\sum_{n=1}^N\left[ \left(\dfrac 3 2 n -\dfrac 3 4 N -1\right)a^\dagger_{k_x,n}a_{k_x,n} \right.\notag \\
&\ \ \ \ \ \ \ \left.+ \left(\dfrac 3 2 n -\dfrac 3 4 N -\dfrac 1 2\right)b^\dagger_{k_x,n}b_{k_x,n}\right],
\end{align}
Then the dielectric polarization is given by 
\begin{align}
P(\varepsilon)=-\dfrac{2}{\sqrt{3}a(N-2/3)}\sum_{n=1}^N\int\dfrac{dk_x}{2\pi}\dfrac{\partial E_n(k_x)}{\partial \varepsilon},
\end{align}
where $E_n(k_x)$ is the $n$th lowest eigenvalue of $H(k_x)$, and $\sqrt{3}a(N-2/3)$ is the  width  of the ribbon. We set the Fermi energy to be $E_f=0$. The calculation results are shown in Fig. \ref{fig:fig5-6}(f). It shows that when $t_2>2t_1$, the dielectric polarization takes a value $P(\varepsilon)\sim\pm e/(2a)$ even for a very weak electric field. This indicates that a half of an electron per unit cell of an edge is accumulated at the edge whenever $t_2>2t_1$. Thus, the anomalous dielectric response occur in this system when it is insulating.

%%%FIG7%%%%
   \begin{figure}[tbp]
    \centering
    \includegraphics[width=86mm]{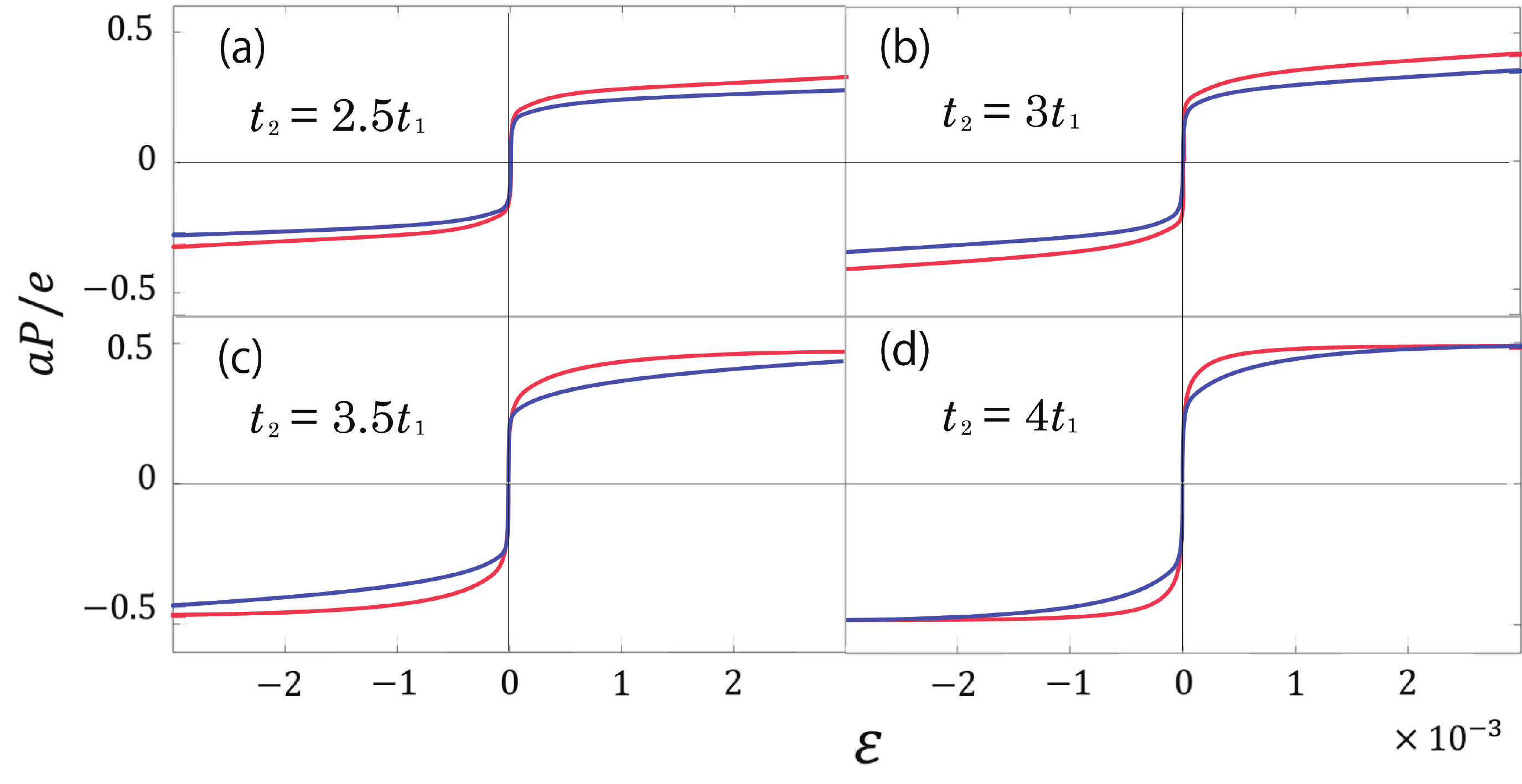}
    \caption{Dielectric polarization of the Model III (red lines) and that of the effective two-site  model (blue lines). We set $N=10$ for the Model II. The hopping amplitudes are taken as $t_1=1$, (a) $t_2=2.5$, (b) $t_2=3$, (c) $t_2=3.5$ and (d) $t_2=4$.}
    \label{fig:EffPolHon}
   \end{figure}
%%%%%%%%%%%

We attribute this phenomenon to topological edge states. To confirm this, we use an effective model which describes the present model at a fixed value of $k_x$, similar to Sec.~II. It consists of two sites, each of which corresponds to one of the edge sites of the ribbon. We call the two sites $L$ and $R$, and the state with the electron occupying the R[L] site is represented as $\ket{R}=(1,0)^T \left [\ket{L}=(0,1)^T ]\right)$ (see Fig. \ref{fig:fig5-6}(b)). Then, the effective Hamiltonian is given by
\begin{align}
H^{\ast}(k_x)=\left(\begin{array}{cc}
{\sqrt{3}(N-2/3)a\varepsilon}/{4} & -t^\ast(k_x)\\
-t^\ast(k_x) &-{\sqrt{3}(N-2/3)a\varepsilon}/{4}
\end{array}\right)\ ,\label{NEHL:efH}
\end{align}
where $t^\ast(k_x)$ is the effective hopping amplitude.
Since $t^\ast(k_x)$ depends on the wave vector $k_x$, it cannot be determined in the same way as in the previous section. Instead, we analytically calculate the edge states of the model, by applying the method in Ref.~\cite{wakabayashi2010electronic} to our anisotropic model. The edge states when $t_2>2t_1$ is given by
\begin{align}
\begin{split}
\ket{\Psi^{1}_{\rm edge}(k_x)}&=C_N\sum_{n=1}^N(-1)^n\sinh[\eta_k(N+1-n)a^\dagger_{k_x,n}\ket{0},\\
\ket{\Psi^{2}_{\rm edge}(k_x)}&=C_N\sum_{n=1}^N(-1)^{N+1-n}\sinh[\eta_kn]a^\dagger_{k_x,n}\ket{0},
\end{split}\label{EdgeStateAniHom}
\end{align}
where 
\begin{align}
|C_N|^2=2[\sinh(\eta_kN)\cosh(\eta_k(N+1))/\sinh(\eta_k)-N]\label{normalization}
\end{align}
is a normalization constant. Here, $\eta_k$ is a non-zero solution of 
\begin{align}
\begin{cases}
-t_1\sinh\eta_{k_x} N-g_k\sinh\eta_{k_x}(N+1)=0,\ \ \text{for}\  0<|ak_x|<\pi,\\
-t_1\sinh\eta_{k_x} N+g_k\sinh\eta_{k_x}(N+1)=0,\ \ \text{for}\  \pi<|ak_x|<2\pi,
\end{cases}
\end{align}
where $g_k=-2t_1\cos\frac{ka}{2}$, and we assume $t_2>2t_1$ and $N\gg1$. The derivation of this equation is shown in Appendex \ref{appA}. One can evaluate the effective hopping amplitude $t^{\ast}(k_x)$ as
%\begin{align}
%t_2\sinh(\eta_k N)\pm 2t_1 \cos\dfrac{k_xa}{2} \sinh[\eta_k(N+1)]=0\label{ConditionOfp}
%\end{align}
%Eq.~(\ref{ConditionOfp}) always has two non-zero solutions for all $k_x$ when $t_2>2t_1$. The derivation of (\ref{EdgeStateAniHom}) is shown in (Appendix. \ref{appA}). Using the above results, the effective hopping amplitude is given by 
\begin{align}
&t^\ast(k_x)\notag \\
&=-\bra{L(k_x)}H_0(k_x)\ket{R(k)}\notag\\
&=(-1)^{N+1}2N|C_N|^2\left[t_1\left(\cosh(\eta_kN)-\dfrac{\sinh(\eta_kN)}{\sinh(\eta_k)}\right)\right. \notag \\
&\left.+t_2\cos\dfrac {k_x} 2\left(\cosh(\eta_kN)-\dfrac{\sinh(\eta_k(N+1))}{\sinh(\eta_k)}+1\right)\right].
\end{align}
The effective dielectric polarization is given by
\begin{align}
P^\ast(\varepsilon)&=\int_0^{2\pi/a}\dfrac{dk_x}{2\pi}p^{\ast}(\varepsilon,k_x)\\
p^\ast(\varepsilon,k_x)&=\frac{e}{2}\left[\braket{L|-}\braket{-|L}-\braket{R|-}\braket{-|R}\right]\notag\\
&=\text{sgn}(\varepsilon)\dfrac{e}{2\sqrt{(2t^\ast(k_x)/(3N/2-1)ae\varepsilon)^2+1}},
\end{align}
where $\ket{-}$ is the eigenstate of (\ref{NEHL:efH}) at $k_x$ with a negative eigenvalue. We plot the polarization of this effective model in Fig.~\ref{fig:EffPolHon}. It agrees well with the result from the Model II, meaning that our scenario of the anomalous polarization is valid. 

%%%
\subsection{Model IV: Model with three orbitals per lattice site on the honeycomb lattice}\label{MTOL}
%%%
Next, we consider a tight-binding model with three orbitals, $s,$ $p_x $ and $p_y$, per lattice site as shown in Fig. \ref{fig:fig5-6} (d). The Hamiltonian is given by
\begin{align}
H_0=-\sum_{\langle ij\rangle}\sum_{\alpha,\beta=s,x,y}t^{ij}_{\alpha\beta}c^{\dagger}_{i\alpha}c_{j\beta}\ ,\label{HamilOfThreeHon}
\end{align}
where $s$, $x$ and $y$ represent $s,p_x$ and $p_y$ orbitals respectively. $c^{\dagger}_{i\alpha}$ and $c_{i\alpha}$ are creation and annihilation operators for an electron in the $\alpha$ orbital at the $i$ site. The hopping integral $t^{ij}_{\alpha\beta}$ is determined from the Slater-Koster parameters \cite{slater1954simplified} $V_{ss\sigma}, V_{sp\sigma}, V_{pp\sigma}$ and $V_{pp\pi}$ as shown in Table \ref{SKElemnts}. In the same way as in the previous section, we can obtain the bulk Hamiltonian $H^{bulk}(\bm k)$ in a $6\times6$ matrix form. Using its eigenstates, we numerically confirmed that the gap is open around $E=0$ and the Zak phase along the $y$-axis becomes $\pi$ in the whole Brillouin zone when $(V_{ss\sigma}, V_{sp_\sigma}, V_{pp\sigma}, V_{pp\pi})=(4.43, -3.79, -5.66, 1.83)$. These values are taken from those for carbon atoms in diamond obtained by fitting from the values of the first-principle calculation  \cite{laref1998calculation} in the unit of eV. We set the Fermi energy to be $E_f=0$. Therefore, an anomalous dielectric response is expected with these  Slater-Koster parameters.
 
 %%%TABLE1%%%
   \begin{table}[tb]
   \caption{Expression of  $t^{ij}_{\alpha\beta}$ for the model IV and the model V \cite{hattori2017edge}. The table shows the hopping amplitude $t^{ij}_{\alpha\beta}$ from the $\alpha$ orbital at the site $i$ to the $\beta$ orbital at the site $j~(\alpha,\beta=s,p_x,p_y,p_z).$ Here $l_{ij}$, $m_{ij}$ and $n_{ij}$ are  the $x, y$ and $z$ components of   the direction cosines measured from site $i$ to $j$. Other components such as $t^{sy}_{ij}, t^{yy}_{ij}$ and $t^{yz}_{ij}$ are obtained from the components in the table by a replacement with the correspondence, $x\leftrightarrow l_{ij}$, $y\leftrightarrow m_{ij}$, and $z\leftrightarrow n_{ij}$.}
   \begin{tabular}{c|l}
   \multicolumn{2}{c}{$t^{ij}_{\alpha\beta}$} \ \\ \hline
   \ $t_{ij}^{ss}$ \  &\ \ \ $l_{ij}V_{ss}$\ \\
   \ $t_{ij}^{sx}=-t_{ij}^{xs}$ \  &\ \ \ $l_{ij}V_{sp}$\ \\
   \ $t_{ij}^{xx}$ \  &\ \ \  $l^2_{ij}V_{pp\sigma}+(1-l^2_{ij})V_{pp\pi}$\ \\
   \ $t_{ij}^{xy}$ \  &\ \ \ $l_{ij}m_{ij}(V_{pp\sigma}-V_{pp\pi})$\ \\
   \end{tabular}\label{SKElemnts}
   \end{table}
%%%%%%%%

%%%FIG8%%%%
   \begin{figure}[tbp]
    \centering
    \includegraphics[width=86mm]{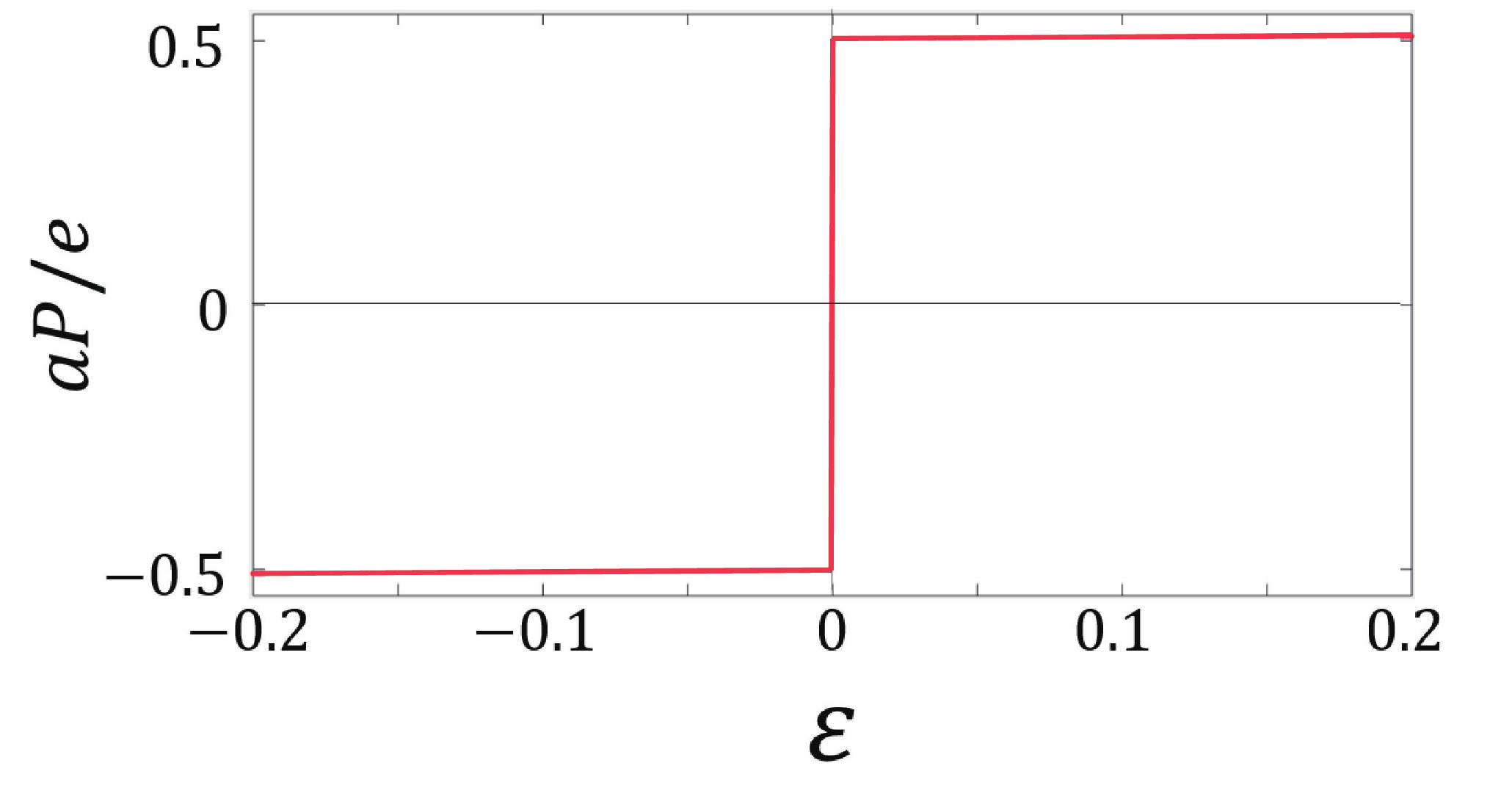}
    \caption{Dielectric polarization of the Model IV in a zigzag-edge nano ribbon. The model is on the honeycomb lattice with anisotropic hopping amplitudes. $P(\varepsilon)$ rapidly changes toward $\pm e/(2a)$ in the vicinity of $\varepsilon=0$. Here we set $N=10$ and the number of sites in the unit cell of the ribbon is $20$. The Slater-Koster parameters $(V_{ss},V_{sp},V_{pp\sigma},V_{pp\pi})$ =  $(4.43,-3.79,-5.66,1.83)$.}
    \label{fig:PolTOHR}
   \end{figure}
%%%%%%%%%%%
The Hamiltonian of this system is given by
\begin{align}
H(k)&=H_0(k)+H_\varepsilon(k),\label{HamilThreeHon}\\
H_\varepsilon&=\dfrac{ae\varepsilon}{\sqrt{3}}\sum_{\alpha}\sum_{n=1}^N\left[ \left(\dfrac 3 2 n -\dfrac 3 4 N -1\right)a^\dagger_{k,\alpha,n}a_{k,\alpha,n} \right.\notag \\
&\ \ \ \ \ \ \ \left.+ \left(\dfrac 3 2 n -\dfrac 3 4 N -\dfrac 1 2\right)b^\dagger_{k,\alpha,n}b_{k,\alpha,n}\right],
\end{align}
where $a_{k,\alpha,n}^\dagger(b_{k,\alpha,n}^\dagger)$ is the creation operator of an electron  with the Bloch wave vector $k$ in the $\alpha$ orbital at the $n$th A(B) site. $H_0(k)$ is the Bloch form of the Hamiltonian $H_0$ with the Bloch wavevector $k$ along the $x$ axis. The dielectric polarization is given by 
\begin{align}
P(\varepsilon)=-\dfrac{2}{\sqrt{3}a(N-2/3)}\sum_{n=1}^{3N}\int_0^{2\pi/a}\dfrac{dk}{2\pi}\dfrac{\partial E_n(k)}{\partial \varepsilon}
\end{align}
where $E_n(k)$ is the $n$th lowest eigenvalue of (\ref{HamilThreeHon}). The calculation results are shown in Fig. \ref{fig:PolTOHR}. The dielectric polarization takes a value $P(\varepsilon)\sim\pm e/(2a)$ for a very weak electric field. Thus, an anomalous dielectric response occurs  even in this case, and it is also attributed to midgap edge states as is similar to the models I and II.

%AAAAA

\section{Analytical derivation of the edge states for Model III}\label{appA}
In this section, we show the derivation of the eigenstates of (\ref{HamRibAn}). In Ref.~\cite{wakabayashi2010electronic}, wavefunctions for the edge states of Model II in an isotropic case is analytically derived. Here we extend this method to an anisotropic case. According to Ref.~\cite{wakabayashi2010electronic}, eigenstates of the model on the anisotropic honeycomb lattice in a nano-ribbon are expressed as 
\begin{align}
\ket{\Psi(k_x)}=\sum_{n=1}^{N}\left[a_{n}(k_x)\ket{A_n(k_x)}+b_{n}(k_x)\ket{B_n(k_x)}\right],\label{EigenstateAni}
\end{align}
where $k_x$ is the wave number along the edge, and $a_n(k_x) [b_n(k_x)]$ denotes the coefficient for the wave function at $A_n [B_n]$ site, $\ket{A_n(k_x)} [\ket{B_n(k_x)}]$.
By using Eqs. (\ref{HamRibAn}) and (\ref{EigenstateAni}), the matrix elements of the Hamiltonian are given by 
\begin{align}
\begin{split}
\bra{B_n(k_x)}H_0(k_x)\ket{A_n(k_x)}&=-2t_1\cos\dfrac{ka}{2}\equiv g_k\\
\bra{B_{n-1}(k_x)}H_0(k_x)\ket{A_n(k_x)}&=-t_2.
\end{split}
\end{align}
Here, we impose a boundary condition 
\begin{align}
a_{N+1}(k_x)=b_0(k_x)=0.
\end{align}
The Schr$\ddot{\text o}$dinger equation $H_0(k_x)\ket{\Psi(k_x)}=E_{k_x}\ket{\Psi(k_x)}$ gives  the following equations;
\begin{align}
\begin{split}
&g_ka_{n}(k_x)-t_2a_{n+1}(k_x)=E_{k_x}b_{n}(k_x),\\
&g_kb_{n}(k_x)-t_2b_{n-1}(k_x)=E_{k_x}a_{n}(k_x).\\
\end{split}\label{app1}
\end{align}
Because it can be expressed as an eigenvalue problem of a $2N \times 2N$ matrix, it has $2N$ solutions. We assume the following forms for $a_n(k_x)$ and $b_n(k_x)$:
\begin{align}
\begin{split}
a_n(k_x)&=Ae^{ipan}+Be^{-ipan},\\
b_n(k_x)&=Ce^{ipan}+De^{-ipan}.
\end{split}
\end{align} 
From the boundary condition, the coefficients $B$ and $D$ are determined as $B=-Az^2$ and $D=-C$ with $z=e^{ip(N+1)}$. Substituting these values into Eq.~(\ref{app1}), we get 
\begin{align}
M\begin{pmatrix}
A\\
C
\end{pmatrix}
=0,\ M=\begin{pmatrix}
m_{11} & m_{12}\\
m_{21}& m_{22}
\end{pmatrix},\label{app2}
\end{align}
where the matrix elements are
\begin{align}
\begin{split}
m_{11}&=-E_{k_x}(e^{ipan}-z^2e^{-ipan}),\\
m_{12}&=e^{ipan}(g_k-t_1e^{-ipa})-e^{-ipan}(g_k-t_1e^{ipa}),\\
m_{21}&=e^{ipan}(g_k-t_1e^{ipa})-e^{-ipan}z^2(g_k-t_1e^{-ipa}),\\
m_{22}&= -E_{k_x}(e^{ipan}-e^{-ipan}).
\end{split}
\end{align}

Equation~(\ref{app2}) has nontrivial solutions only when $\det M=0$. Then, we get the following two equations;
\begin{align}
E_{k_x}^2&=(g_k-t_2e^{ipa})(g_k-t_2e^{-ipa}),\label{appE}\\
E_{k_x}^2(1+z^2)&=(g_k-t_2e^{-ipa})^2z^2+(g_k-t_2e^{ipa})^2.\label{app3}
\end{align}
By eliminating $E_ {k_x}$ from these equations, we obtain an equation for $p$:
\begin{align}
F(p,N)\equiv -t_2\sin paN+g_k\sin pa(N+1)=0.\label{app6}
\end{align}
This equation holds when $pa = 0,\pm\pi$. However, this corresponds to the trivial solutions $a_n = b_n = 0$ for all $n$, and should be discarded. As shown in Ref.~\cite{wakabayashi2010electronic}, there are only $2N-2$ solutions which correspond to the bulk states for $p$ when
\begin{align}
%-\dfrac{N}{N+1}t_2<g_k<\dfrac{N}{N+1}t_2,\label{app4}
|g_k|\le\dfrac{N}{N+1}t_2.\label{app4}
\end{align}
 The remaining two solutions are expected to correspond to the edge states. These edge states can be obtained by analytical continuation as
\begin{align}
pa=\begin{cases}
&\pi\pm i\eta_{k_x}=p_\pi,\ k^L_ca<|k_xa|<\pi, \\ 
&0\pm i\eta_{k_x}=p_0,\ \pi<|k_xa|<k^R_ca,
\end{cases}\label{appp}
\end{align}
where the wavevectors $k^L_c$ and $k^R_c$ are defined as 
\begin{align}
\begin{cases}
\cos \dfrac{k^L_c a}{2}&=\dfrac{t_2}{2t_1}\dfrac{N}{N+1},\\
\cos \dfrac{k^R_c a}{2}&=-\dfrac{t_2}{2t_1}\dfrac{N}{N+1}.
\end{cases}
\end{align}
Substituting (\ref{appp}) into (\ref{app6}), we get
\begin{align}
\begin{cases}
-t_1\sinh\eta_{k_x} N-g_k\sinh\eta_{k_x}(N+1)=0,\ \ \text{for}\  p_{\pi},\\
-t_1\sinh\eta_{k_x} N+g_k\sinh\eta_{k_x}(N+1)=0,\ \ \text{for}\  p_{0}.
\end{cases}\label{app5}
\end{align}
There exists one solution for $\eta_{k_x}$ each equation in Eq.~(\ref{app5}) when Eq. (\ref{app4}) is satisfied. When $t_2>2t_1$ and in the limit of large $N$, Eq.~(\ref{app5}) is always satisfied and the two edge states are given by Eq.~(\ref{appp}).

Since we focus on the edge states, we consider only the solution of (\ref{app5}). Then, the coefficients $a_n$ and $b_n$ are given by
\begin{align}
a_n(k_x)&=\pm(-1)^{N+1-n}C_N\sinh\eta_{k_x}(N+1-n),\\
b_n(k_x)&=(-1)^{n}C_N\sinh\eta_{k_x}n,
\end{align}
where $C_N$ is the normalization factor expressed in (\ref{normalization}). The two edge states $\ket{\Psi_{\rm edge}^+(k_x)}$ and $\ket{\Psi_{\rm edge}^-(k_x)}$ are obtained by substituting these into (\ref{EigenstateAni}). These edge states $\ket{\Psi_{\rm edge}^\pm(k_x)}$ are not localized at one edge, but are symmetric and antisymmetric linear combinations of edge states at an either end. Therefore, to obtain approximate edge states at one edge, we take linear combinations as follows:
\begin{align}
\ket{\Psi^{1}_{\rm edge}(k_x)}&=\left(\ket{\Psi_{edge}^+(k_x)}+\ket{\Psi_{edge}^-(k_x)}\right)/\sqrt{2},\\
\ket{\Psi^{2}_{\rm edge}(k_x)}&=\left(\ket{\Psi_{edge}^+(k_x)}-\ket{\Psi_{edge}^-(k_x)}\right)/\sqrt{2}.
\end{align}
These are given in Eq. (\ref{EdgeStateAniHom}) and represent states localized at one end.
%\begin{acknowledgments}
%\end{acknowledgments}


%%%
\section{Three-dimensional Model V: Model with four orbitals per lattice site on the diamond lattice}\label{DLFO}
\begin{figure}[tbp]
    \centering
    \includegraphics[width=86mm]{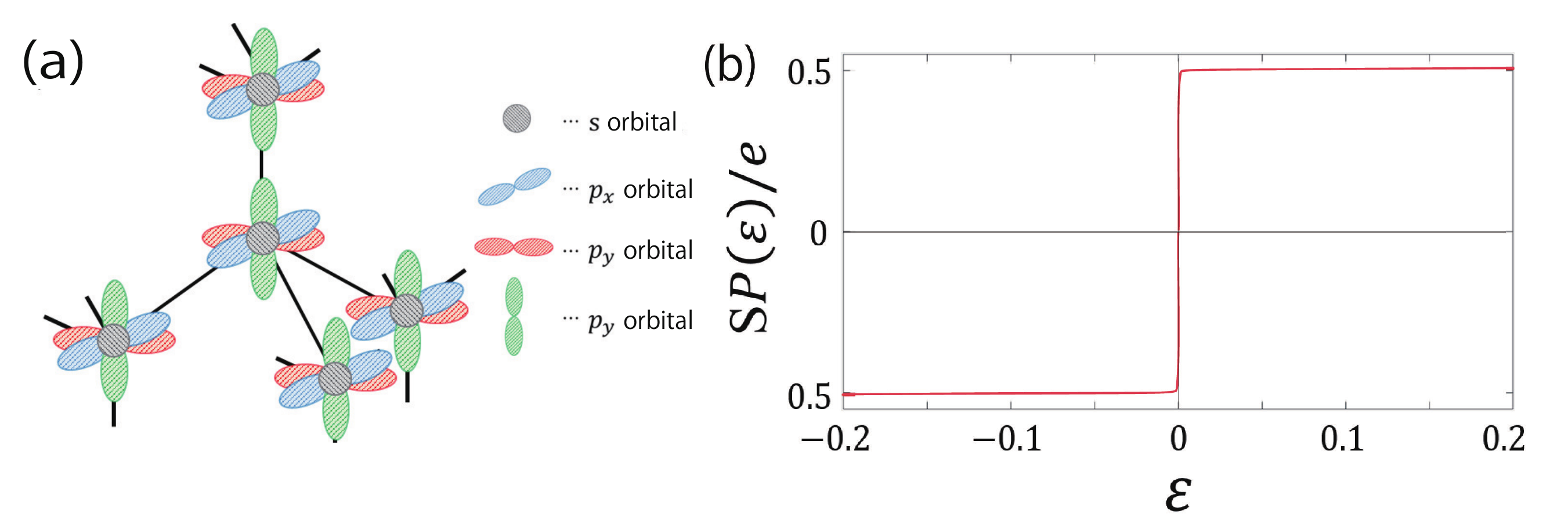}
    \caption{Model V on the diamond lattice. (a) Schematic figure of the model with four orbitals $s, p_x, p_y$ and $p_z$ orbitals per lattice site (Model V). (b) Dielectric polarization of  a slab of the Model V on a diamond lattice with the (111) surfaces. $P(\varepsilon)$ rapidly changes toward $\pm e/(2S)$ in the vicinity of $\varepsilon=0$, where $S$ is an area of the surface unit cell. Here we set $N=10$ and the number of sites in the unit cell is $20$. The Slater-Koster parameters are $(V_{ss},V_{sp},V_{pp\sigma},V_{pp\pi})$ =  $(4.43,-3.79,-5.66,1.83)$.}
    \label{fig:ModelV}
\end{figure}

%%%
In this section, we show the anomalous dielectric response in a three-dimensional tight-binding model (Model V) on the diamond lattice, with four orbitals per unit cell. As is similar to Model II, we use the diamond lattice and its slab with (111) surfaces are shown in Fig. \ref{fig:ModelII} (a) and (b). Then, we consider a tight-binding model with four orbitals, $s, p_x, p_y$ and $p_z$, per site on the diamond lattice as shown in Fig. \ref{fig:ModelV} (d). The Hamiltonian is 
\begin{align}
H_0=-\sum_{\langle ij\rangle}\sum_{\alpha,\beta=s,x,y,z}t^{ij}_{\alpha\beta}c^{\dagger}_{i\alpha}c_{j\beta}\ ,\label{HamilOfFourHon}
\end{align}
where $s$, $x$, $y$ and $z$ represent $s,p_x$,$p_y$ and $p_z$ orbitals, respectively. Here, $c^{\dagger}_{i\alpha}$ and $c_{i\alpha}$ are creation and annihilation operators for an electron with the $\alpha$ orbital at $i$ site. The hopping integrals $t^{ij}_{\alpha\beta}$ are taken as the Slater-Koster form shown in Table \ref{SKElemnts}. We can obtain the bulk Hamiltonian $H^{bulk}(\bm k)$ as an $8\times8$ matrix form. Using its eigenstates, we numerically confirmed that the gap is open and that the Zak phase along the $z$-axis becomes $\pi$ in the whole Brillouin zone when $(V_{ss\sigma}, V_{sp_\sigma}, V_{pp\sigma}, V_{pp\pi})=(4.43, -3.79, -5.66, 1.83)$, which are obtained by fitting from the values of the first-principle calculation with the unit of eV \cite{laref1998calculation}. We take the Fermi energy to be $E_f=0$. Therefore, the anomalous dielectric response is expected when the Slater-Koster parameters take these values.

The Hamiltonian of the slab with the (111) surfaces is given by
\begin{align}
H(\bm k_\parallel)&=H_0(\bm k_\parallel)+H_\varepsilon(\bm k_\parallel),\label{HamFourOrbSlabElec}\\
H_\varepsilon(\bm k_\parallel)&=\notag\\
\dfrac{\sqrt{6}}{4}a&e\varepsilon\sum_{n=1}^{N}\sum_{\alpha=s,x,y,z}\left[ \left(\dfrac 4 3 n -\dfrac 2 3 N -\dfrac 5 6\right)a^\dagger_{\bm k_\parallel,\alpha,n}a_{\bm k_\parallel,\alpha,n} \right.\notag \\
&\ \ \ \ \ \ \ \left.+ \left(\dfrac 4 3 n -\dfrac 2 3 N -\dfrac 1 2\right)b^\dagger_{\bm k_\parallel,\alpha,n}b_{\bm k_\parallel,\alpha,n}\right],
\end{align}
where $a_{\bm k_\parallel,\alpha,n}^\dagger(b_{\bm k_\parallel,\alpha,n}^\dagger)$ is a creation operator of an electron with a Bloch wave vector $\bm k_\parallel$ in the $\alpha$ orbital of the $n$th A(B) site in the unit cell.  The dielectric polarization is given by 
\begin{align}
P(\varepsilon)=-\dfrac{1}{a(4N/3-1)}\sum_{n=1}^{4N}\int_{\bm k_\parallel\in \text{2DBZ}}\dfrac{d^2\bm k_\parallel}{(2\pi)^2}\dfrac{\partial E_n(\bm k_\parallel)}{\partial \varepsilon},
\end{align}
where $E_n(\bm k_\parallel)$ is the $n$th lowest eigenvalue of (\ref{HamFourOrbSlabElec}). The calculation results are shown in  Fig. \ref{fig:ModelV} (f). The dielectric polarization takes a value $P(\varepsilon)=\pm e/(2S)$ even for a very weak electric field. This indicates that a half of an electrons are accumulated at the surface per unit cell. Thus, the anomalous dielectric response occurs in this system when it is insulating.

\section{Computational conditions}\label{appB}

The electronic structures of
%diamond and silicon
diamond, silicon, carbon nanotube, and PTFE
are obtained from the GGA of the DFT.
We use the ab initio code OpenMX based on localized
basis functions and norm-conserving pseudopotentials.
We use the Perdew-Burke-Ernzerhof (PBE) functional in the GGA~\cite{Perdew96}.
We employ the $8 \times 8 \times 8$ and $8 \times 8 \times 1$ $\bm{k}$-point mesh for the bulk and the slab calculation for diamond and silicon, respectively,
and the $1 \times 1 \times 16$ $\bm{k}$-point mesh for carbon nanotube and PTFE.
%We take 60 atoms on the (111) surface.
The valence orbital set is 
$s2p2d1$ for C, $s2p3d1$ for Si, and $s2p2d1$ for F.
The energy cutoff for the numerical integrations is 150 Ry.

%\bibliography{cite}


\end{document}